\documentclass[12pt,a4paper]{article}
\usepackage{slashed,cite}
\usepackage{amssymb}
\usepackage{amsmath}
\usepackage{amsfonts,latexsym}
\usepackage{bbm}
\usepackage{color,epsfig}

\def\be{\begin{equation}}
\def\ee{\end{equation}}
\def\bea{\begin{eqnarray}}
\def\eea{\end{eqnarray}}
\def\nn{\nonumber}

\csname @addtoreset\endcsname{equation}{section}
\newcommand{\ft}[2]{{\textstyle\frac{#1}{#2}}}

\newsavebox{\uuunit}
\sbox{\uuunit}
    {\setlength{\unitlength}{0.825em}
     \begin{picture}(0.6,0.7)
        \thinlines
        \put(0,0){\line(1,0){0.5}}
        \put(0.15,0){\line(0,1){0.7}}
        \put(0.35,0){\line(0,1){0.8}}
       \multiput(0.3,0.8)(-0.04,-0.02){12}{\rule{0.5pt}{0.5pt}}
     \end {picture}}

\def\d{\delta}

\def\e{\epsilon}

\def\l{\lambda}

\def\m{\mu}
\def\n{\nu}

\newcommand{\id}{\mathbbm{1}}

\begin{document}
\begin{titlepage}

\begin{flushright}
  UG-02-41 \\
  hep-th/0209205
\end{flushright}
\vskip 1.5truecm

\begin{center}
\vspace{.3cm} \baselineskip=16pt
{\LARGE \bf (Non-)Abelian Gauged Supergravities \\ ~\\ in
  Nine Dimensions}
 \vskip 2truecm

{\large E.~Bergshoeff, T.~de~Wit, U.~Gran, R.~Linares and D.~Roest} \\
\vskip 6 mm {\small
   Centre for Theoretical Physics, University of Groningen,\\
   Nijenborgh 4, 9747 AG Groningen, The Netherlands. \\
   E-mail: {\tt \symbol{`\{}e.a.bergshoeff, t.c.de.wit, u.gran, r.linares, d.roest\symbol{`\}}@phys.rug.nl}}
\end{center}
\vskip 2truecm

\centerline{ABSTRACT}
\bigskip

We construct five different two--parameter massive deformations of the unique nine--dimensional $N=2$
supergravity. All of these deformations have a
higher--dimensional
origin via Scherk--Schwarz reduction and correspond to gauged supergravities.
The gauge groups we encounter are $SO(2)$, $SO(1,1)^+$, ${\mathbb R}$, ${\mathbb R}^+$ and the two--dimensional
non--Abelian Lie group A(1), which consists of scalings and
translations in one dimension.

We make a systematic search for half-supersymmetric domain walls
and non-supersym\-metric de Sitter space solutions.
Furthermore, we discuss in which sense the supergravities we have constructed can be considered as
low-energy limits of compactified superstring theory.

\end{titlepage}

\section{Introduction}

It is well-known that the low-energy limit of superstring theory and/or
M--theory is described by a supergravity theory in the same spacetime dimension
and with the same number of supersymmetries. Thus, M--theory leads to
D=11 supergravity and Type IIA/IIB superstring theory leads to D=10 IIA/IIB
supergravity. The same applies to the compactifications of these
theories to lower dimensions.

The other way round is less clear: not
every supergravity theory has necessarily a string or M--theory
origin.
A well-known example of a supergravity theory whose role
in string theory was unclear until a few years ago
is the D=10 massive supergravity theory of Romans \cite{Romans:1986tz}.
It was pointed out by Romans
that the D=10 (massless) IIA supergravity theory
\cite{Campbell:1984zc,Giani:1984wc}
can be
deformed into a massive supergravity with mass parameter $m_{\rm R}$.
The role of this massive supergravity within string theory
has become clear only after the introduction of the D--branes, in particular
the D8--brane \cite{Polchinski:1995mt}.
An interesting feature of the massive supergravity of Romans is that
the Lagrangian possesses a dilaton potential proportional to $m_{\rm R}^2$
which acts as an effective cosmological constant. Due to this scalar
potential the
massive supergravity, unlike the massless case, does not
allow a maximally supersymmetric Minkowski spacetime as a vacuum
solution. Instead, the scalar potential leads to the
possibility of a half-supersymmetric
domain wall solution interpolating between different
values of the cosmological constant. Such a solution indeed
exists \cite{Polchinski:1996df,Bergshoeff:1996ui} and is identified as the
D8-brane of \cite{Polchinski:1995mt}.

The massive supergravity of \cite{Romans:1986tz}
is {\it not} a gauged supergravity and, at the field theory level,
has {\it no} D=11 origin\footnote{We assume that we are not using
the existence of extra Killing vectors, like in \cite{Bergshoeff:1998ak}.}.
The only candidate symmetry of the Lagrangian to be gauged is a rigid
${\mathbb R}^+$ symmetry (see Table~\ref{IIA_weights}).
However, the Ramond-Ramond gauge vector has a
nontrivial weight under this ${\mathbb R}^+$ symmetry and this leads to
inconsistencies with the supersymmetry algebra.
There does exist another massive deformation
of D=10 IIA supergravity, with mass parameter $m_{11}$,
which {\it is} a gauged supergravity and {\it does} have
a D=11 origin \cite{Howe:1998qt,Lavrinenko:1998qa}. However, it can only
be defined at the level of the equations of motion.
The Ramond-Ramond gauge
vector has weight zero with respect to the ${\mathbb R}^+$ group that is gauged
(see Table~\ref{IIA_weights}) and in this case there are no inconsistencies
with the supersymmetry algebra.
The role of this second massive deformation within string theory
is not (yet) clear. An interesting feature of the theory is that
it allows for a (non-supersymmetric)
de Sitter space solution \cite{Lavrinenko:1998qa}. The
possible physical significance of this de Sitter space solution has
been discussed in \cite{Chamblin:2001dx,Chamblin:2001jj}.

A common feature of the D=10 massive supergravity of \cite{Romans:1986tz}
and the D=10 gauged supergravity of \cite{Howe:1998qt,Lavrinenko:1998qa}
is that there is a dilaton potential which is proportional
to the square of the mass parameter, $m_{\rm R}^2$ and $m_{11}^2$,
respectively.
Due to this scalar potential
the D=10 Minkowski spacetime is no longer a maximally
supersymmetric vacuum solution of the theory. Instead one can look for
half-supersymmetric vacuum solutions. A natural class of
half-supersymmetric solutions that makes use of the scalar
potential is the set of domain-wall solutions, like the D8--brane
mentioned above.
Recently, domain wall solutions of lower-dimensional supergravities
have attracted attention in view of their relevance for
a supersymmetric Randall-Sundrum scenario \cite{Randall:1999ee,Randall:1999vf},
the domain-wall/QFT correspondence \cite{Boonstra:1998mp,Behrndt:1999mk}
and applications to cosmology \cite{Kallosh:2001gr,Townsend:2001ea}.
In all these applications the properties of the domain walls
play a crucial role and these properties are determined by the
details of the scalar potential.

Motivated by this we studied in a previous paper general domain wall
solutions in D=9 dimensions \cite{Bergshoeff:2002mb}{}\footnote{
For earlier discussions of domain wall solutions in D=9 dimensions,
see \cite{Cowdall:2000sq,Gheerardyn:2001jj}. For a more recent discussion, see 
\cite{Nishino:2002zi}.}.
We took D=9 because on the one hand this
case shares some of the complexities of the lower-dimensional cases, on the
other hand the scalar potential for this case is simple enough to study
the corresponding domain-wall solutions in full detail.
The supergravity theory we considered in \cite{Bergshoeff:2002mb}
was obtained by a
generalized Scherk-Schwarz reduction of D=10 IIB supergravity. This is
not the most general possibility in D=9.
The aim of this paper is to make a systematic search for massive
deformations of the unique D=9, N=2 supergravity theory.
All deformations we find correspond to {\sl gauged} supergravities.
Such supergravities have a gauge symmetry which
reduces, for constant values of the gauge parameter, to a nontrivial
rigid symmetry. The hope is that the D=9 case will
teach us something about the more complicated situation in
${\rm D}<9$ dimensions.

In the first part of this paper
we will present in two steps the D=9 gauged supergravities
we have found. In a first step we will present seven massive deformations
with a {\it single}
mass parameter $m$, all giving rise to gauged supergravities.
All of them are obtained by generalized dimensional reduction
\cite{Scherk:1979ta} from a higher-dimensional theory (11D, IIA or
IIB supergravity). The consistency of these 9D gauged supergravities is
guaranteed by their higher--dimensional origin. The gauge groups
we encounter are either\footnote{Throughout the paper we will use the notation $SO(1,1)^+$ rather than (the isomorphic) ${\mathbb R}^+$ for the scaling symmetry that is a subgroup of $SL(2,\mathbb{R})$. The different notation is used to emphasize the different origin.} $SO(2)$,
$SO(1,1)^+$, ${\mathbb R}$ (all of which are subgroups of $SL(2,\mathbb{R})$
with invariant metrics diag\,(1,1), diag\,(1,--1) and diag\,(1,0), respectively
\cite{Hull:1998vy,Hull:2002wg}), ${\mathbb R}^+$ or the two--dimensional
non--Abelian Lie group A(1)\footnote{This is unrelated to the A-D-E
classification of simple Lie groups.}. The latter is the affine group of the line and
consists of so-called collinear transformations (scalings and
translations) in one dimension and forms a non--semi--simple Lie group \cite{Frankel, Gilmore}.

In a second step we will consider combinations of these seven massive
deformations. The closure of the supersymmetry algebra will be guaranteed
based on a linearity argument but it turns out that non-linear restrictions
enter via the back door. Satisfying these restrictions leaves us with five
different two--parameter deformations (rather than the seven--parameter
deformation that one could have if there were no non-linear restrictions).
These are the most general gauged supergravities we construct in
this paper.

In the second part of this paper we make a systematic search for
vacuum solutions of the supergravities we have
obtained. The existence of such vacuum solutions is needed in
order to define the spectrum of the theory as fluctuations around this vacuum.
Our search includes half-supersymmetric domain wall solutions
and non-supersymmetric de Sitter spaces.

Throughout the paper we will reduce field equations rather than Lagrangians.
The reason for this is the fact that (some of) the rigid symmetries we employ 
for Scherk--Schwarz reduction scale the Lagrangian. As was noted by
\cite{Lavrinenko:1998qa} and is illustrated by the SS reduction of
a simple toy model in Appendix~\ref{appendix:toymodel},
Scherk--Schwarz reduction with a symmetry that scales the Lagrangian can
only be performed at the level of the field equations.
Reduction of the Lagrangian itself gives rise to the wrong equations.
In fact, the reduced field equations can not be obtained as Euler--Lagrange
equations of {\it any} Lagrangian.

This paper is organized as follows. In Section~\ref{section:11D}
we briefly review the situation in D=11 dimensions
where no massive deformation has
been constructed so far. This case is needed for the
discussion of the D=10 and D=9 cases.
In Section~\ref{section:IIA} we discuss the two different
massive deformations of D=10 IIA supergravity.
In Section~\ref{section:IIB} we review the case of D=10 IIB supergravity.
This case does not allow massive deformations but will be needed for the
discussion of the D=9 case. In Section~\ref{section:9D} we present 7
massive deformations of the maximally
supersymmetric D=9 supergravity theory. They all are gauged
supergravities with gauge group $SO(2)$, $SO(1,1)^+$, ${\mathbb R}$, ${\mathbb R}^+$ or A(1).
In Section~\ref{section:combinations} we show, by combining
the different gauged supergravities, that there exist five different
two--parameter massive supergravity theories.
In Section~\ref{section:solutions} we make a systematic search for
vacuum solutions of the supergravities we have
obtained. Finally, in Section~\ref{section:conclusions} we give our
conclusions. In particular, we discuss which of the
D=9 supergravities can be considered as candidate
low-energy limits of (compactified) superstring theory.
We give four Appendices. Appendix~\ref{appendix:conventions}
contains our conventions. Appendix~\ref{appendix:toymodel} discusses the
Scherk--Schwarz reduction of a dilaton--gravity toy model.
Appendix~\ref{appendix:Ansatze} contains the
supersymmetry transformations of {\it massless}
D=11, 10 and 9 supergravity plus the reduction Ans\"atze to go
from D=11 to D=10 to D=9. Finally, in Appendix~\ref{appendix:spinors}
we discuss some manipulations with spinors and gamma-matrices in
ten and nine dimensions.

\section{D=11 Supergravity} \label{section:11D}

We first consider eleven-dimensional supergravity.
Its field content is given by\footnote{
In order to distinguish between D=11, D=10 and D=9 we indicate D=11
fields and indices with a double hat, D=10 fields and indices with a single hat
and D=9 fields and indices with no hat.}
\begin{align}
   \text{D=11:} \qquad \{ \hat{\hat e}_{\hat{\hat\mu}}{}^{\hat{\hat a}} ,
\hat{\hat C}_{\hat{\hat\mu}\hat{\hat\nu}\hat{\hat\rho}} ,
\hat {\hat \psi}_{\hat {\hat \mu}} \} \,.
\end{align}

\noindent
The D=11 supersymmetry transformations are given in
Appendix~\ref{appendix:Ansatze}, see eq.~\eqref{11Dsusy}.
These supersymmetry rules are covariant under an ${\mathbb R}^+$ symmetry
with parameter $\hat{\hat \alpha}$ \cite{Bergshoeff:1996cg}.
The weights of the D=11 fields under this
${\mathbb R}^+$ are given in Table~\ref{11d_weights}.
Note that the Lagrangian is not invariant but scales with weight $w=9$. Therefore this ${\mathbb R}^+$ is a symmetry of the equations of motion only.

\begin{table}[ht]
\begin{center}
\begin{tabular}{||c||c|c|c|c||c||}
\hline \rule[-2.5mm]{0mm}{8mm}
 ${\mathbb R}^+$ & $\hat{\hat e}_{\hat{\hat\mu}}{}^{\hat{\hat a}}$ &
$\hat{C}_{\hat{\hat\mu}\hat{\hat\nu}\hat{\hat\rho}}$ &
$\hat{\hat \psi}_{\hat{\hat\mu}}$ & $\hat{\hat\e}$ &
$\hat{\hat{\mathcal{L}}}$ \\
\hline \hline \rule[-1mm]{0mm}{6mm}
$\hat{\hat \alpha}$ & $1$ & $3$ & $\tfrac{1}{2}$ & $\tfrac{1}{2}$ &
$9$ \\
\hline
\end{tabular}
\caption{\it The ${\mathbb R}^+$--weights of the D=11 supergravity fields,
the supersymmetry parameters $\hat {\hat \epsilon}$ and the Lagrangian
$\hat {\hat {\cal {L}}}$.}
\label{11d_weights}
\end{center}
\end{table}

No massive deformation of the eleven-dimensional supergravity theory is known.
In particular, no cosmological constant can be added \cite{Nicolai:1981kb}.
One problem with a D=11 supersymmetric cosmological constant is that its
reduction gives rise to a D=10 cosmological constant with a dilaton
coupling that differs from Romans' massive deformation.
A general deformation of D=11 supergravity involving the use of extra Killing
vectors has been considered in \cite{Meessen:1998qm}.
We will not consider this possibility in this paper.

\section{Massive Deformations of D=10 IIA Supergravity} \label{section:IIA}

A Kaluza-Klein reduction of the eleven-dimensional theory
yields the IIA theory in ten dimensions\footnote{
  We have used the reduction Ans\"atze~\eqref{11Dred} with $m_{11}=0$. }.
The field content of the D=10 IIA supergravity theory is given by
\begin{align}
  \text{D=10 IIA:} \qquad \{ \hat{e}_{\hat\mu}{}^{\hat a}, \hat{B}_{\hat\mu\hat\nu}, \hat{\phi}, \hat A_{\hat\mu}, \hat{C}_{\hat\mu\hat\nu\hat\rho}, \hat{\psi}_{\hat\mu}, \hat{\lambda} \}\, .
\end{align}
The supersymmetry transformations rules are given in
eq.~\eqref{IIAsusy}. For later purposes we indicate these (undeformed)
supersymmetry transformations by $\delta_0$.
The transformation rules have two ${\mathbb R}^+$--symmetries,
one with parameter $\hat\alpha$ that scales the Lagrangian and one
with parameter $\hat\beta$ that leaves the Lagrangian invariant.
The first symmetry follows via dimensional reduction from the
D=11 ${\mathbb R}^+$--symmetry with parameter $\hat{\hat \alpha}$.
The weights of these two ${\mathbb R}^+$--symmetries are given in Table~\ref{IIA_weights}.
The gauge symmetry associated to the Ramond-Ramond vector, with parameter
$\hat\lambda$, reads
\begin{align}
{\hat A} \rightarrow {\hat A} - {\rm d} \hat \lambda \,, \qquad
  {\hat C} \rightarrow {\hat C} - {\rm d} \hat\lambda \, {\hat B} \,.
\label{gtrRR}
\end{align}

\begin{table}[ht]
\begin{center}
\begin{tabular}{||c||c|c|c|c|c|c|c|c||c||c||}
\hline \rule[-1mm]{0mm}{6mm}
 ${\mathbb R}^+$ & $\hat{e}_{\hat\mu}{}^{\hat a}$ & $\hat{B}_{\hat\mu\hat\nu}$ &
$e^{\hat{\phi}}$ & $\hat A_{\hat\mu}$ &
$\hat{C}_{\hat\mu\hat\nu\hat\rho}$ & $\hat{\psi}_{\hat\mu}$ & $\hat{\lambda}$
& $\hat \epsilon$ & $\hat{\mathcal{L}}$ & Origin \\
\hline \hline \rule[-2mm]{0mm}{6mm}
$\hat{\alpha}$ &
$\ft{9}{8}$ &
$3$ &
$\ft{3}{2}$ &
$0$ &
$3$ &
$\ft{9}{16}$ &
$-\ft{9}{16}$ &
$\ft{9}{16}$ &
$9$ & $\hat {\hat \alpha}$ \\
\hline \rule[-2mm]{0mm}{6mm}
$\hat{\beta}$ & 0 &
$\ft12$ &
1 &
$-\ft34$ &
$-\ft14$ &
$0$ &
$0$ &
$0$ &
$0$ & \\
\hline
\end{tabular}
\caption{\it The ${\mathbb R}^+$--weights of the D=10 IIA supergravity fields,
the supersymmetry parameter $\hat\epsilon$ and the Lagrangian $\hat {\cal L}$.}
\label{IIA_weights}
\end{center}
\end{table}

The D=10 IIA supergravity theory allows two massive deformations which
we discuss one by one below.

\subsection{Deformation $m_{\rm R}$: D=10 massive supergravity}
\label{subsection:10d_gauged_sugra}

The first massive deformation, with mass parameter $m_{\rm R}$,
is due to Romans \cite{Romans:1986tz}.
In this case (the same is true for all other cases)
the supersymmetry transformations receive two types of massive deformations:
explicit and implicit ones. The explicit deformations are terms, at most linear
in $m_{\rm R}$, that are added to the original supersymmetry rules.
These explicit deformations are denoted by $\delta_{m_{\rm R}}$ and
are given, in terms of a superpotential
$W(\hat\phi)$ and derivatives thereof, by
\begin{align}
  m_{\rm R}:
  \begin{cases}
  \delta_{m_{\rm R}} {\hat \psi} _{\hat \mu}
  = - \ft{1}{8} W \hat{\Gamma}_{\hat \mu} \hat \epsilon \,, \qquad
  \text{with~~} W = \ft{1}{4} e^{5 \hat \phi /4} m_{\rm R} \,, \\
  \delta_{m_{\rm R}} \hat \lambda = 4 \ft{\delta W}{\delta \hat \phi} \hat \epsilon \,.
  \end{cases}
\end{align}
There are further implicit massive deformations to the original
supersymmetry rules $\delta_0$, which are given in eq.~\eqref{IIAsusy},
 due to the fact that in these rules one must replace all field strengths by corresponding {\it massive} field strengths which are given by
\begin{align}
  {\hat F} & = {\rm d} {\hat A} + m_{\rm R} {\hat B} \,, \qquad
  {\hat H} = {\rm d} {\hat B} \,, \qquad
  {\hat G} = {\rm d} {\hat C} + {\hat A} {\hat H} + \ft{1}{2} m_{\rm R}
{\hat B} {\hat B} \,.
\end{align}
The Lagrangian contains terms linear and quadratic in $m_{\rm R}$.
Again there are implicit deformations, via the massive field strengths,
and explicit deformations. The explicit deformations quadratic in the
mass parameter define the scalar potential which can be written
in terms of the superpotential
$W(\hat\phi)$ and derivatives thereof.

Requiring closure of the supersymmetry algebra one finds the
linear deformations of the fermionic (gravitino and dilatino)
field equations in Roman's theory:
\begin{align}
  m_{\text{R}}:
  \begin{cases}
  {{X}}_{m_{\text{R}}} (\hat{\psi}^{\hat{\mu}}) \equiv
  m_{\text{R}} e^{5 \hat{\phi} /4} \hat{{\Gamma}}^{\hat{{\mu}}\hat{{\nu}}}
  (\tfrac{1}{4} \hat{{\psi}}_{\hat{{\nu}}} +
  \tfrac{5}{288} \hat{{\Gamma}}_{\hat{{\nu}}} \hat{\lambda}) \,, \\
  {{X}}_{m_{\text{R}}} (\hat{\lambda}) \equiv
  m_{\text{R}} e^{5 \hat{\phi} /4} \hat{{\Gamma}}^{\hat{{\nu}}}
  (-\tfrac{5}{4} \hat{{\psi}}_{\hat{{\nu}}} -
  \tfrac{21}{160} \hat{{\Gamma}}_{\hat{{\nu}}} \hat{\lambda}) \,.
  \end{cases}
\end{align}
The undeformed equations,
${{X}}_0 (\hat{\psi}^{\hat{\mu}})$ and
${{X}}_0 (\hat{\lambda})$, are given in eqs.~(\ref{X0IIA}).

Under supersymmetry the fermionic field equations,
$X_0 + {{X}}_{m_{\text{R}}}$, transform into
the deformed bosonic equations of motion. Since we will only be
interested in finding solutions that are carried by the
metric and the scalars it is convenient to truncate away all
bosonic fields except the metric and the dilaton\footnote{Note
that a further truncation to $\phi = c $ is inconsistent.}.
After this truncation we find that under supersymmetry the fermionic
field equations transform into
\begin{align}
  (\delta_0+\delta_{m_{\text{R}}}) &
  ({{X}}_0 + {{X}}_{m_{\text{R}}}) (\hat{\psi}^{\hat{\mu}})
  = \tfrac{1}{2} \hat{{\Gamma}}^{\hat{{\nu}}} \hat{{\epsilon}} \,
  [\hat{{R}}^{\hat{{\mu}}}{}_{\hat{{\nu}}}
  - \tfrac{1}{2} \hat{{R}} \hat{{g}}^{\hat{{\mu}}}{}_{\hat{{\nu}}}
  - \tfrac{1}{2} (\partial^{\hat{{\mu}}} \hat{\phi}) (\partial_{\hat{{\nu}}} \hat{\phi})
  + \tfrac{1}{4} (\partial \hat{\phi})^2 \hat{{g}}^{\hat{{\mu}}}{}_{\hat{{\nu}}}
  + \tfrac{1}{4} m_{\text{R}}^2 e^{5 \hat{\phi}/2} \hat{{g}}^{\hat{{\mu}}}{}_{\hat{{\nu}}}]
  \,, \notag \\
  (\delta_0+\delta_{m_{\text{R}}}) & ({{X}}_0 + {{X}}_{m_{\text{R}}}) (\hat \lambda) =
  \hat{\epsilon} \, [ \Box \hat{\phi} - \tfrac{5}{4} m_{\text{R}}^2 e^{5 \hat{\phi}/2} ]\,.
\end{align}
At the right-hand side we thus find the
Romans' bosonic field equations for the metric and the dilaton,
one solution of which is the D8-brane. Note that the bosonic field
equations contain terms quadratic in the mass parameter.

Romans' theory is not known to have a higher-dimensional supergravity origin.
Neither is it a gauged supergravity. A candidate symmetry of the Lagrangian
to be gauged is the $\hat\beta$ symmetry of Table~\ref{IIA_weights}.
However, the candidate gauge field $\hat {A}_{\hat\mu}$ has a nontrivial
weight under $\hat \beta$.
This means that the curl $d\hat A$ transforms with a non-covariant term
proportional to $d\hat\lambda\hat A$. Such a term cannot be cancelled
by adding an extra term, such as $\hat B$,
 to the definition of the $\hat A$ curvature.
In short, the $\hat\beta$ symmetry cannot be gauged
\cite{Giani:1984wc}.
The same Table shows that on the other hand $\hat {A}_{\hat\mu}$ has
weight zero under the $\hat\alpha$--symmetry which is a symmetry of the
equations of motion only. This $\hat\alpha$--symmetry can indeed be gauged
at the level of the equations of motion. This gauging leads to the
D=10 gauged supergravity discussed below.

\subsection{Deformation $m_{11}$: D=10 gauged supergravity}

The second massive deformation, with mass parameter $m_{11}$,
has been considered in \cite{Howe:1998qt, Lavrinenko:1998qa} and is a
gauged supergravity.
It can be obtained by generalized Scherk-Schwarz reduction of D=11
 supergravity using the ${\mathbb R}^+$ symmetry $\hat{\hat{\alpha}}$
of Table~\ref{11d_weights} \cite{Lavrinenko:1998qa}.
The corresponding reduction Ans\"atze, with
$m_{11} \ne 0$, are given in eq.~\eqref{11Dred}. This reduction leads to the
following explicit massive deformations of the D=10 IIA supersymmetry rules:
\begin{align}
m_{11}:
\begin{cases}
  \delta_{m_{11}} {\hat \psi} _{\hat \mu}
    = \ft{9}{16} m_{11} e^{-3\hat \phi/4} \hat{\Gamma}_{\hat \mu}
  \Gamma_{11} \hat \epsilon \,, \\
  \delta_{m_{11}} \hat \lambda = \ft{3}{2} m_{11} e^{-3\hat \phi/4}
  \Gamma_{11} \hat \epsilon \,.
\end{cases}
\end{align}
The implicit massive deformations of the original supersymmetry rules
$\delta_0$ are given by the massive bosonic field strengths
\begin{align}
  {\rm D}{\hat\phi} & = {\rm d}{\hat\phi} + \ft32 m_{11} {\hat A} \,, \qquad
  {\hat F} = {\rm d} {\hat A} \,, \qquad
  {\hat H} = {\rm d} \hat B + 3 m_{11} \hat C \,, \qquad
  {\hat G} = {\rm d} {\hat C} + {\hat A} {\hat H} \,,
\end{align}
while the covariant derivative of the supersymmetry parameter is given by
\begin{align}
  D_{\hat \mu} \hat \epsilon & = (\partial_{\hat \mu} +
\hat{\omega}_{\hat \mu} + \ft{9}{16} m_{11} \hat{\Gamma}_{\hat \mu}
\slashed{\hat{A}} ) \hat \epsilon \,.
\end{align}
The gauge vector in the definition of the covariant derivative is required to 
make the
derivative of the supersymmetry parameter {\it and} the spin connection
${\mathbb R}^+$--covariant.

The linear deformations of the fermionic field equations read in this case
\begin{align}
  m_{11}:
  \begin{cases}
  {{X}}_{m_{11}} (\hat{\psi}^{\hat{\mu}}) \equiv
  m_{11} e^{-3 \hat{\phi} /4} \Gamma_{11} \hat{{\Gamma}}^{\hat{{\mu}}\hat{{\nu}}}
  (-\tfrac{9}{2} \hat{{\psi}}_{\hat{{\nu}}} +
  \tfrac{17}{48} \hat{{\Gamma}}_{\hat{{\nu}}} \hat{\lambda}) \,, \\
  {{X}}_{m_{11}} (\hat{\lambda}) \equiv
  m_{11} e^{-3 \hat{\phi} /4} \Gamma_{11} \hat{{\Gamma}}^{\hat{{\nu}}}
  (\tfrac{3}{2} \hat{{\psi}}_{\hat{{\nu}}} -
  \tfrac{9}{16} \hat{{\Gamma}}_{\hat{{\nu}}} \hat{\lambda}) \,.
  \end{cases}
\end{align}
We first consider the truncation that all
bosonic fields except the metric and the dilaton are set equal to zero.
Under supersymmetry the fermionic field equations transform into
\begin{align}
  (\delta_0+\delta_{m_{11}}) ({{X}}_0 + {{X}}_{m_{11}}) (\hat{\psi}^{\hat{\mu}})=
  & \, \tfrac{1}{2} \hat{{\Gamma}}^{\hat{{\nu}}} \hat{{\epsilon}} \,
  \big[ \hat{{R}}^{\hat{{\mu}}}{}_{\hat{{\nu}}}
  - \tfrac{1}{2} \hat{{R}} \hat{{g}}^{\hat{{\mu}}}{}_{\hat{{\nu}}}
  - \tfrac{1}{2} (\partial^{\hat{{\mu}}} \hat{\phi}) (\partial_{\hat{{\nu}}} \hat{\phi})
  + \tfrac{1}{4} (\partial \hat{\phi})^2 \hat{{g}}^{\hat{{\mu}}}{}_{\hat{{\nu}}} + \notag \\
  & \hspace{1cm} + 36 m_{11}^2 e^{-3 \hat{\phi}/2} \hat{{g}}^{\hat{{\mu}}}{}_{\hat{{\nu}}} \big] +
  \Gamma_{11} \hat{\epsilon}
  [ 3 m_{11} e^{-3 \hat{\phi}/4} \partial^{\hat{\mu}} \hat{\phi} ] \,, \notag \\
  (\delta_0+\delta_{m_{11}}) ({{X}}_0 + {{X}}_{m_{11}}) (\hat \lambda) =
  & \, \hat{\epsilon} \, [ \Box \hat{\phi} ] +
  \hat{{\Gamma}}^{\hat{{\nu}}} \Gamma_{11} \hat{\epsilon}
  [ 9 m_{11} e^{-3 \hat{\phi}/4} \partial_{\hat{\nu}} \hat{\phi} ] \,.
\end{align}
The terms involving $\Gamma_{11}$ are part of the vector field equation.
Therefore, to obtain a consistent truncation, we must further
truncate the dilaton to zero.
One is then left with only the metric satisfying the Einstein equation with
a positive cosmological constant, a solution of which is 10D de Sitter space
\cite{Lavrinenko:1998qa}.

The reduced theory is a gauged supergravity where the ${\mathbb R}^+$ symmetry 
$\hat \alpha$ of Table~\ref{IIA_weights} has been gauged. In particular,
the gauge parameter
and transformation of the Ramond-Ramond potentials read as follows{}\footnote{
It is understood that each field with $w_{\hat\alpha}\ne 0$ is multiplied
by $\Lambda$.}:
\begin{align}
  \hat \alpha: \qquad
  \Lambda = e^{w_{\hat \alpha} m_{11} \hat \lambda} \qquad \text{with} \qquad
  {\hat A} \rightarrow {\hat A} - {\rm d} \hat \lambda \,, \qquad
  {\hat C} \rightarrow e^{3 m_{11} \hat \lambda} ({\hat C} - {\rm d} \hat \lambda
 \, {\hat B}) \,,
\end{align}
where $w_{\hat \alpha}$ are the weights under $\hat \alpha$.
We note that one can take two different limits of the $\hat\alpha$ gauge
transformations. First, the limit $m_{11} \rightarrow 0$ leads to
the massless gauge transformations \eqref{gtrRR}.
Note that $\hat C$ transforms trivially under this gauge symmetry in the sense
that $\hat C$ can be made gauge-invariant after a simple field-redefinition.
Secondly, one can take the limit that $\hat\alpha$ is constant. This leads to
the ungauged ${\mathbb R}^+$ $\hat\alpha$--symmetry of Table~\ref{IIA_weights}.

A noteworthy feature of the D=10 gauged supergravity is that no Lagrangian
can be defined for it. In the search for supersymmetric domain wall
solutions in D=5 dimensions many other examples of gauged supergravity
theories without a Lagrangian have been given \cite{Bergshoeff:2002qk}.
Note that one can write down a Lagrangian for the ungauged
theory. The reason that one cannot write down a
Lagrangian after gauging is that the symmetry that is gauged
is not a symmetry of the Lagrangian but only of the equations
of motion. It would be instructive to construct the D=10 gauged supergravity
from the ungauged theory by gauging the $\hat\alpha$--symmetry.
Apparently, it shows that one can gauge symmetries that leave
a Lagrangian invariant up to a scale factor.

\section{D=10 IIB Supergravity} \label{section:IIB}

The other ten-dimensional supergravity theory is chiral IIB.
Its field content is
\begin{align}
  \text{D=10 IIB:} \qquad \{ \hat{e}_{\hat\mu}{}^{\hat a}, \hat{\phi}, \hat \chi, \hat{B}^{(1)}_{\hat\mu\hat\nu}, \hat{B}^{(2)}_{\hat\mu\hat\nu}, \hat{D}_{\hat\mu\hat\nu\hat\rho\hat\sigma}, \hat{\psi}_{\hat\mu}, \hat{\lambda} \}\, .
\label{sb}
\end{align}
The supersymmetry variations are given in eq.~\eqref{IIBsusy}.
The IIB supersymmetry rules transform covariant under the $SL(2,\mathbb{R})$ transformations (omitting indices):
\begin{align}
  \hat{\tau} & \rightarrow \frac{a \hat{\tau} +b}{c\hat{\tau} +d} \,, \qquad
  \vec{\hat{B}} \rightarrow \Omega \vec{\hat{B}} \,, \qquad
  \hat D \rightarrow \hat D \,, \qquad
  \text{with~~} \Omega =
    \left( \begin{array}{cc} a&b\\c&d \end{array} \right) \in SL(2,\mathbb{R}) \,,
    \notag \\
  \hat{\psi}_{\hat{\mu}} & \rightarrow
    \left( \frac{c \, \hat{\tau}^*+d}{c\, \hat{\tau}+d} \right)^{1/4}
    \hat{\psi}_{\hat{\mu}} \,, \qquad
  \hat{\lambda} \rightarrow \left( \frac{c \, \hat{\tau}^*+d}{c \, \hat{\tau}+d}
    \right)^{3/4} \hat{\lambda} \,, \qquad
  \hat{\epsilon} \rightarrow \left( \frac{c \, \hat{\tau}^*+d}{c \, \hat{\tau}+d}
    \right)^{1/4} \hat{\epsilon} \,.
\end{align}
We have used here the vector notation $\vec{\hat{B}} = \bigl (\hat {B}^{(1)}, \hat {B}^{(2)}\bigr )^T$.
The group $SL(2,\mathbb{R})$ contains a set of three one-parameter
conjugacy classes defining one compact and two non--compact subgroups. Since they are
needed later we will describe them shortly. Each of the subgroups is generated
by a $SL(2,\mathbb{R})$ group element $\Omega$ with det\,$\Omega = 1$.

\begin{description}

\item{1.}
One non--compact subgroup ${\mathbb R}$ is generated
by
\begin{equation}
\Omega_p = e^{\frac{1}{2}\hat\zeta\, (\sigma_1 + i\sigma_2)} =
 \left( \begin{array}{cc} 1&\hat\zeta\\0&1 \end{array} \right)\, .
\end{equation}
Each element defines a parabolic conjugacy class with Tr\,$\Omega =2$.
These parabolic transformations leave the combination
$(\hat {B}^{(2)})^2$ invariant. Therefore the invariant metric is given by
diag\,(0,1). The action of the ${\mathbb R}$ $\hat\zeta$--symmetry on the
fields can not be expressed by assigning weights to the standard basis
of fields given in \eqref{sb}.

\item{2.} An $SO(1,1)^+$ subgroup which is generated by elements
\begin{equation}
\Omega_h = e^{\hat\gamma\,\sigma_3} =
  \left( \begin{array}{cc} e^{\hat\gamma}&0\\
0&e^{-\hat\gamma} \end{array} \right)\, .
\end{equation}
Each element defines a hyperbolic conjugacy class with Tr\,$\Omega >2$.
These hyperbolic transformations leave the combination $
\hat {B}^{(1)}\hat {B}^{(2)}$ invariant. After diagonalization this
leads to an invariant metric given by diag\,(1,--1). The weights
corresponding to the $SO(1,1)^+$ $\hat\gamma$--symmetry are given in Table~\ref{IIB_weights}.

\item{3.}
There is a $SO(2)$ subgroup which is generated by elements $\Omega$
of $SL(2,\mathbb{R})$
with
\begin{equation}
 \Omega_e = e^{i\hat\theta\,\sigma_2} =
\left( \begin{array}{cc} {\rm cos}\, \hat\theta& {\rm sin}\, \hat\theta
\\-{\rm sin}\, \hat\theta& {\rm cos}\, \hat\theta \end{array} \right)\, .
\end{equation}
Each element defines an elliptic conjugacy class with Tr\,$\Omega<2$.
The elliptic transformations leave $(\hat {B}^{(1)})^2 + (\hat {B}^{(2)})^2$
invariant. After diagonalization this leads to an invariant metric
given by diag\,(1,1). The action of the $SO(2)$ $\hat\theta$--symmetry on the
fields can not be expressed by assigning weights to the standard real
basis of fields given in \eqref{sb}.

\end{description}

\noindent
Table 3 contains the weights of the $\hat\gamma$--symmetry defined
above\footnote{The other two symmetries defined above cannot
 be defined in terms of weights of real fields only.}
and of a new ${\mathbb R}^+$ symmetry $\hat \delta$
which is {\it not} a subgroup of $SL(2,\mathbb{R})$ and that does not leave
the Lagrangian invariant.
One could combine $SL(2,\mathbb{R})$ with this new ${\mathbb R}^+$
into a $GL(2,\mathbb{R})$ symmetry that leaves the IIB equations of motion
invariant.
Its action is the product of the two separate transformations: $\tilde \Omega = \Omega \Lambda_{\hat \delta}$.
This exhausts all the symmetries of D=10 IIB
supergravity.

\begin{table}[h]
\begin{center}
\begin{tabular}{||c||c|c|c|c|c|c|c|c|c||c||c||}
\hline \rule[-1mm]{0mm}{6mm}
$\mathbb{R}^+$ & $\hat{e}_{\hat\mu}{}^{\hat a}$ &
$e^{\hat{\phi}}$ & $\hat \chi$ & $\hat{B}^{(1)}_{\hat\mu\hat\nu}$ & $\hat{B}^{(2)}_{\hat\mu\hat\nu}$ &
$\hat{D}_{\hat\mu\hat\nu\hat\rho\hat\sigma}$ & $\hat{\psi}_{\hat\mu}$ & $\hat{\lambda}$
& $\hat \epsilon$ & $\hat{\mathcal{L}}$ & symmetry \\
\hline
\hline \rule[-2mm]{0mm}{6mm}
$\hat{\gamma}$ &
$0$ &
$-2$ &
$2$ &
$1$ &
$-1$ &
$0$ &
$0$ &
$0$ &
$0$ &
$0$ &
$SO(1,1)^+$ \\
\hline \rule[-2mm]{0mm}{6mm}
$\hat{\delta}$ &
$1$ &
$0$ &
$0$ &
$2$ &
$2$ &
$4$ &
$\tfrac{1}{2}$ &
$-\tfrac{1}{2}$ &
$\tfrac{1}{2}$ &
$8$ & ${\mathbb R}^+$\\
\hline
\end{tabular}
\caption{\it The scaling weights of the D=10 IIB supergravity fields,
the supersymmetry parameter $\hat\epsilon$ and the Lagrangian $\hat {\cal L}$.}
\label{IIB_weights}
\end{center}
\end{table}

The IIB supergravity theory is not known to have massive deformations.
One of the reasons for this is that there is no candidate vector field
like in the IIA case.

\section{Massive deformations of D=9, $N=2$ Supergravity} \label{section:9D}

The Kaluza-Klein reduction of either (massless) IIA or IIB supergravity gives the unique $D=9$, $N=2$ massless supergravity theory. Its field content is given by
\begin{align}
  \text{D=9:} \qquad \{ e_\mu{}^a, \phi, \varphi, \chi, A_\mu, A^{(1)}_\mu, A^{(2)}_\mu, B^{(1)}_{\mu\nu}, B^{(2)}_{\mu\nu}, C_{\mu\nu\rho}, \psi_\mu, \lambda, \tilde \lambda \}\, .
\end{align}
The supersymmetry rules are given in eq.~\eqref{9Dsusy}.
The massless 9--dimensional theory inherits several global symmetries from
its parents: two ${\mathbb R}^+$ symmetries $\alpha,\beta$ from IIA supergravity
and one ${\mathbb R}^+$ symmetry $\delta$ plus a full $SL(2,\mathbb{R})$ symmetry
from IIB supergravity. The latter leads in particular to an $SO(2)$
symmetry $\theta$, an $SO(1,1)^+$ symmetry $\gamma$ and an
${\mathbb R}$--symmetry $\zeta$. The weights of all
these symmetries, except for the $SO(2)$ $\theta$--symmetry and
${\mathbb R}$ $\zeta$--symmetry, and their higher-dimensional origin are given 
in Table~\ref{9d_weights} (see also \cite{Bergshoeff:1996cg}).

\begin{table}[ht]
\hspace{-1cm}
\begin{tabular}{||c||c|c|c|c|c|c|c|c|c|c|c|c|c|c||c||c||}
\hline \rule[-1mm]{0mm}{6mm}
  $\mathbb{R}^+$ & $e_\mu{}^a$ & $e^\phi$ & $e^\varphi$ & $\chi$ & $A_\mu$ &
  $A^{(1)}_\mu$ & $A^{(2)}_\mu$ & $B^{(1)}_{\mu\nu}$ & $B^{(2)}_{\mu\nu}$ &
  $C_{\mu\nu\rho}$ & $\psi_\mu$ & $\lambda$ & $\tilde \lambda$ & $\epsilon$ & $\mathcal{L}$
  & Origin \\
\hline \hline \rule[-1mm]{0mm}{6mm}
  $\alpha$ & $\tfrac{9}{7}$ & $0$
   & $\tfrac{6}{\sqrt{7}}$ & $0$ & $3$ & $0$ & $0$ &
  $3$ & $3$ & $3$ & $\tfrac{9}{14}$ & $-\tfrac{9}{14}$ & $-\tfrac{9}{14}$ & $\tfrac{9}{14}$ &
  $9$ & 11D \\
\hline \rule[-1mm]{0mm}{6mm}
  $\beta$ & $0$ &
  $\tfrac{3}{4}$ & $\tfrac{\sqrt{7}}{4}$ & -$\tfrac{3}{4}$ & $\tfrac{1}{2}$ & $-\tfrac{3}{4}$ & $0$ & $-\tfrac{1}{4}$ & $\tfrac{1}{2}$ & $-\tfrac{1}{4}$ & $0$ & $0$ & $0$ & $0$ &
  $0$ & IIA \\
\hline \rule[-1mm]{0mm}{6mm}
  $\gamma$ & $0$ &
  $-2$ & $0$ & $2$ & $0$ & $1$ & $-1$ & $1$ & $-1$ & $0$ & $0$ & $0$ & $0$ & $0$ &
  $0$ & IIB \\
\hline \rule[-1mm]{0mm}{6mm}
  $\delta$ & $\tfrac{8}{7}$ &
  $0$ & $-\tfrac{4}{\sqrt{7}}$ & $0$ & $0$ & $2$ & $2$ & $2$ & $2$ & $4$ & $\tfrac{4}{7}$ & $-\tfrac{4}{7}$ & $-\tfrac{4}{7}$ & $\tfrac{4}{7}$ &
  $8$ & IIB \\
\hline
\end{tabular}
\caption{\it The scaling weights of the 9 dimensional supergravity fields, the supersymmetry parameter $\epsilon$ and the Lagrangian $\mathcal{L}$.}
\label{9d_weights}
\end{table}

It turns out that only three out of the four scalings given in Table~\ref{9d_weights} are linearly independent. There is a relation
\begin{equation}
\ft49 \alpha - \ft83 \beta = \gamma + \ft12 \delta\, .
\label{rel}
\end{equation}
We observe the following pattern. Using \eqref{rel} to eliminate one of the scaling--symmetries we are left with three independent scaling--symmetries. Each of the three gauge fields $A_\mu, A_\mu^{(1)}, A_\mu^{(2)}$ has weight zero under {\it two} (linear combinations) of these three symmetries: one is
a symmetry of the action, the other is a symmetry of the equations of motion only.

The D=9 $SL(2,\mathbb{R})$ symmetry acts in the following way:
\begin{align}
  {\tau} & \rightarrow \frac{a {\tau} +b}{c{\tau} +d} \,, \qquad
  \vec{{A}} \rightarrow \Omega \vec{{A}} \,, \qquad
  \vec{{B}} \rightarrow \Omega \vec{{B}} \,, \qquad
  \text{with~~} \Omega =
    \left( \begin{array}{cc} a&b\\c&d \end{array} \right) \in SL(2,\mathbb{R}) \,,
    \notag \\
  {\psi}_{{\mu}} & \rightarrow
    \left( \frac{c \, {\tau}^*+d}{c\, {\tau}+d} \right)^{1/4}
    {\psi}_{{\mu}} \,, \qquad
  {\lambda} \rightarrow \left( \frac{c \, {\tau}^*+d}{c \, {\tau}+d}
    \right)^{3/4} {\lambda} \,, \notag \\
  {\tilde \lambda} & \rightarrow \left( \frac{c \, {\tau}^*+d}{c \, {\tau}+d}
    \right)^{-1/4} {\tilde \lambda} \,, \qquad
  {\epsilon} \rightarrow \left( \frac{c \, {\tau}^*+d}{c \, {\tau}+d}
    \right)^{1/4} {\epsilon} \,,
\label{SL2R9D}
\end{align}
while $\varphi$ and $C$ are invariant. We have used a vector
notation for the two vectors and two antisymmetric tensors, like in D=10.
Again one can combine $SL(2,\mathbb{R})$ with an ${\mathbb R}^+$ symmetry to form $GL(2,\mathbb{R})$ with parameter $\tilde \Omega = \Omega \Lambda_{{\mathbb R}^+}$.

In addition to the global symmetries there is a number of local symmetries.
In particular, the gauge transformations of the vectors read
\begin{alignat}{2}
  A^{(1)} & \rightarrow A^{(1)} - {\rm d} \lambda^{(1)} \,, \qquad &
  A^{(2)} & \rightarrow A^{(2)} - {\rm d} \lambda^{(2)} \,, \notag \\
  A & \rightarrow A - {\rm d} \lambda \,, \qquad &
  \vec{B} & \rightarrow \vec{B} - \vec{A} \, {\rm d} \lambda \,.
\end{alignat}

\begin{figure}[tb]
\centerline{\fbox{\epsfig{file=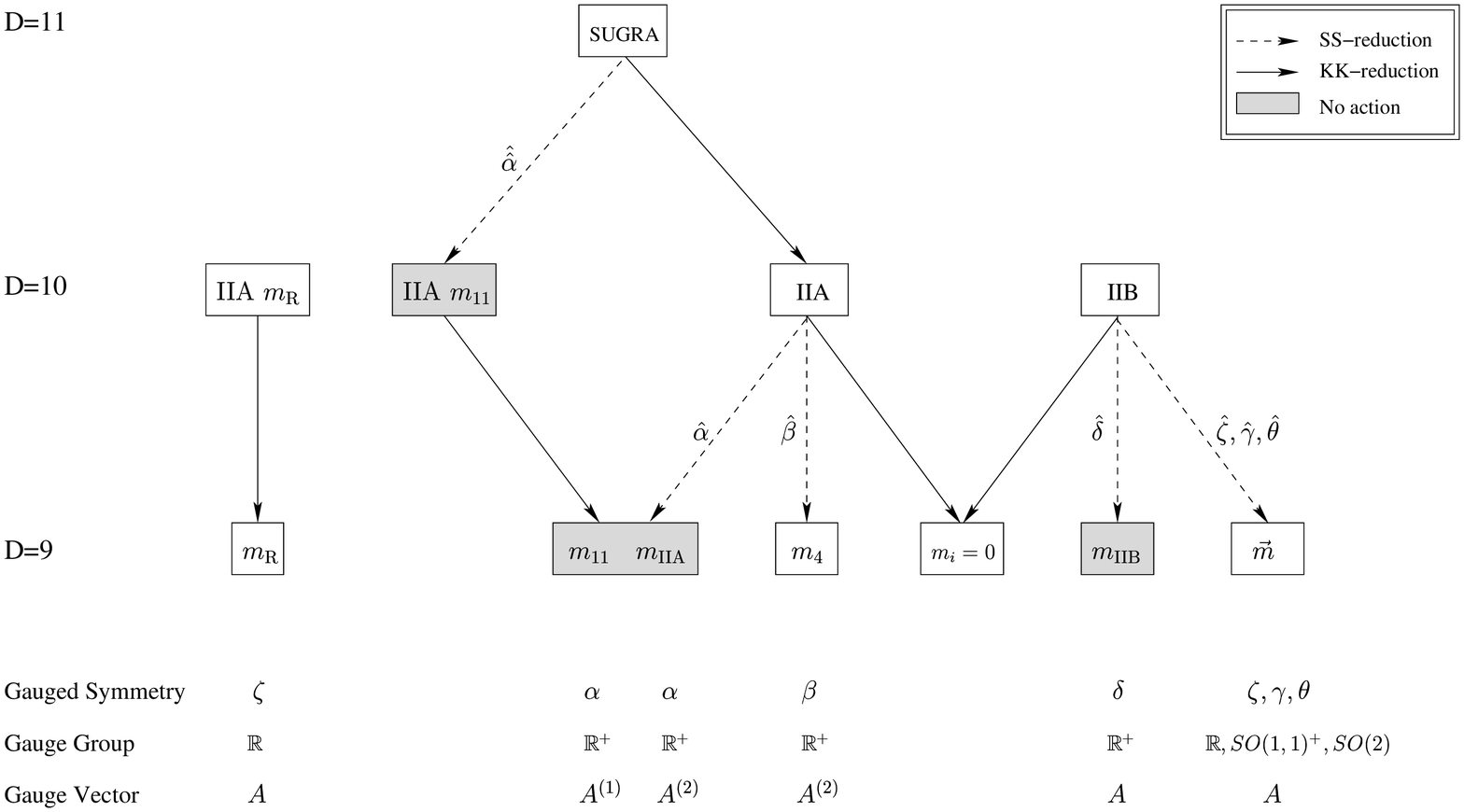,width=.98\textwidth}}}
\caption{\it
Overview of all reductions performed in this paper. These cases can all be interpreted as gauged supergravities, with gauged symmetry and corresponding gauge field as given in the Figure. Mass parameters in the same box, such as $m_{11}, m_{\rm IIA}$ or $m_1, m_2, m_3$, form a multiplet under $SL(2,\mathbb{R})$. Further details of these cases will be given below. Note that the two ways of obtaining the ${\mathbb R}$--gauging give rise to the massive T-duality of~\cite{Bergshoeff:1996ui}.} \label{fig:reductions}
\end{figure}

We now turn to massive deformations of the 9D theory. Applying a SS
dimensional reduction of
the higher-dimensional supergravities we obtain a number of massive deformations in nine dimensions, as illustrated in Figure~\ref{fig:reductions}.
By employing the different global symmetries of 11D, IIA and IIB supergravity we obtain seven deformations of the unique $D=9$ supergravity.

Note that the different massive deformations can be related. Symmetries of the massless theory become field redefinitions in the massive theory that only act on the massive deformations. This means that the mass parameters transform under such transformations: they have a scaling weight under the different scaling symmetries and fall in multiplets of $SL(2,\mathbb{R})$. In Table~\ref{mpar} the multiplet structure of the massive deformations under $SL(2,\mathbb{R})$ is given. The mass parameter $\tilde m_4$ is defined as the S-dual partner of $m_4$ and can not be obtained by a SS reduction of IIA supergravity.

\begin{table}[ht]
\begin{center}
\begin{tabular}{||c||c||}
\hline \rule[-2.5mm]{0mm}{8mm}
 mass parameters & $SL(2,\mathbb{R})$ \\
\hline \hline \rule[-1mm]{0mm}{6mm}
$(m_1,m_2,m_3)$ & triplet \\
$(m_4,\tilde m_4)$ & doublet \\
$(m_{11},m_{\rm IIA})$ & doublet \\
$m_{\rm IIB}$& singlet\\
\hline
\end{tabular}
\caption{\it This tables indicates the different multiplets that
the D=9 mass parameters form under $SL(2,\mathbb{R})$.}
\label{mpar}
\end{center}
\end{table}

All these deformations correspond to a gauging of a 9D global symmetry. In particular, it is always the symmetry that is employed in the SS reduction Ansatz that becomes gauged upon reduction. The corresponding gauge vector is always provided by the metric, i.e. is the Kaluza--Klein vector of the dimensional reduction. In all but one case this is the complete story and one finds an Abelian gauged supergravity. It turns out that there is one exception where
we find a {\it non-Abelian} gauge symmetry.
This can be understood from the following general rule\footnote{We thank Sergio Ferrara for clarifying discussions on this issue.}. As we noted, the Kaluza--Klein vector gauges the symmetry employed in the SS reduction Ansatz. The fate of either of the remaining two gauge vectors is restricted to three possibilities:
\begin{itemize}
\item
The vector is a singlet under the gauge symmetry and its field strength acquires no modification, e.g. $A^{(1)}$ in the $m_{\rm IIA}$ deformation.
\item
The vector transforms under the gauge symmetry and its field strength acquires a massive deformation proportional to a two--form.
The degrees of freedom of the vector are eaten up by the two--form via the St\"uckelberg mechanism, e.g. $A$ in the $m_{\rm IIA}$ deformation.
\item
The vector transforms under the gauge symmetry and its field strength acquires no massive deformation proportional to a two--form.
In this case we must have gauge enhancement to preserve covariance, e.g. $A^{(1)}$ in the $m_4$ deformation.
\end{itemize}
All cases we find in D=9 are consistent with this rule of thumb.
We will discuss the different massive deformations one by one below.

\subsection{Deformation $m_{\rm IIA}$:\, SS Reduction of IIA using $\hat \alpha$}

We first perform a Scherk-Schwarz reduction of (massless) IIA
supergravity based on the
$\hat \alpha$--symmetry of Table~\ref{IIA_weights}. We use the
reduction Ans\"atze \eqref{IIAred} with $m_4=0$.
This leads to a gauged supergravity with mass parameter $m_{\rm IIA}$.
The explicit massive deformations in this case
appear in the variation of the gravitino and one of the dilatinos:
\begin{align}
  m_{\text{IIA}}: &
  \begin{cases}
    \delta_{m_{\text{IIA}}} \psi_\mu = - \tfrac{9}{14} i m_{\text{IIA}}
      e^{\phi/2 - 3\varphi/2\sqrt{7}} \gamma_\mu \epsilon^* \,, \\
    \delta_{m_{\text{IIA}}} \tilde \lambda =
    \tfrac{6}{\sqrt{7}} m_{\text{IIA}} e^{\phi/2 - 3\varphi/2\sqrt{7}} \epsilon \,.
  \end{cases}
\end{align}
The implicit massive deformations are given by
\begin{alignat}{2}
  {\rm D} \phi & = e^{-\phi} {\rm d} e^{\phi} \,, \qquad
  {\rm D} \varphi = e^{-\varphi} {\rm D} e^{\varphi} \,, \qquad
  G_1 = {\rm d} \chi \,, \qquad
  G_4 = {\rm D} C + {\vec B}^T \eta {\vec F} \,, \notag \\
  F & = {\rm D} A - 3 m_{\text{IIA}} B^{(2)} \,, \qquad
  F^{(1)} = {\rm d} A^{(1)} \,, \qquad
  F^{(2)} = {\rm d} A^{(2)} \,, \notag \\
  H^{(1)} & = {\rm D} B^{(1)} - A F^{(1)} +3 m_{\text{IIA}} C \,, \qquad
  H^{(2)} = {\rm D} B^{(2)} - A F^{(2)} \,.
\end{alignat}
The ${\mathbb R}^+$--covariant derivative is defined by
 ${\rm D}={\rm d} + w_\alpha \, m_{\text{IIA}} A^{(2)}$ with $w_\alpha$ the $\alpha$ scaling--weight
of the field it acts on, as given in the Table~\ref{9d_weights}, and
${\rm DD} = w_\alpha \, m_{\text{IIA}} F^{(2)}$. The covariant
derivative of the supersymmetry parameter is given by
\begin{align}
  D_{\mu} \epsilon & = (\partial_{\mu} + {\omega}_{\mu} + \tfrac{i}{4} e^{\phi} \partial_\mu \chi + \ft{9}{14} m_{\text{IIA}} \Gamma_{\mu} \slashed{A}^{(2)} ) \epsilon \,.
\end{align}
The 9D fermionic field equations have the following explicit massive deformations:
\begin{align}
  m_{\text{IIA}}: &
  \begin{cases}
  X_{m_{\text{IIA}}} (\psi^\mu) = i m_{\text{IIA}} e^{\phi/2-3\varphi/2\sqrt{7}} \gamma^{\mu \nu}
    [\tfrac{9}{2}\psi_\nu^* -i\tfrac{9}{32} \gamma_\nu \lambda
    + i \tfrac{3}{4\sqrt{7}} \gamma_\nu \tilde{\lambda} ] \,, \notag \\
  X_{m_{\text{IIA}}} (\lambda) = - m_{\text{IIA}} e^{\phi/2-3\varphi/2\sqrt{7}}
    \gamma^\nu [ - i \tfrac{\sqrt{7}}{6} \gamma_\nu \tilde{\lambda}^* ] \,, \notag \\
  X_{m_{\text{IIA}}} (\tilde{\lambda}) = - m_{\text{IIA}} e^{\phi/2-3\varphi/2\sqrt{7}} \gamma^\nu
    [\tfrac{6}{\sqrt{7}}\psi_\nu -\tfrac{11}{6\sqrt{7}} i \gamma_\nu \lambda^*
    +\tfrac{1}{7} i \gamma_\nu \tilde{\lambda}^* ] \,.
  \end{cases}
\end{align}
This massive deformation is a gauging of the ${\mathbb R}^+$ symmetry $\alpha$:
\begin{align}
   \alpha: \qquad \Lambda = e^{w_\alpha m_{\text{IIA}} \lambda^{(2)}} \qquad \text{with} \qquad
   A^{(2)} \rightarrow A^{(2)} - {\rm d} \lambda^{(2)} \, ,
\end{align}
where $w_\alpha$ are the weights under $\alpha$.

\subsection{Deformation $m_4$: SS reduction of IIA using $\hat\beta$}

We next perform a generalized Scherk-Schwarz reduction of D=10 IIA
supergravity using the ${\mathbb R}^+$ $\hat\beta$ symmetry of Table~\ref{IIA_weights}. We use the reduction Ans\"atze given in eq.~\eqref{IIAred}, taken with $m_{\text{IIA}}=0$. This leads to a massive deformation with mass parameter $m_4$. Only the supersymmetry variations of the dilatinos receive explicit massive deformations:
\begin{align}
  m_4: &
  \begin{cases}
    \delta_{m_4} \lambda =
      \tfrac{3}{4} m_4 e^{\phi/2 - 3\varphi/2\sqrt{7}} \epsilon \,, \\
    \delta_{m_4} \tilde \lambda =
      \tfrac{\sqrt{7}}{4} m_4 e^{\phi/2 - 3\varphi/2\sqrt{7}} \epsilon \,.
  \end{cases}
\label{susyexplmiia}
\end{align}
The implicit massive deformations read:
\begin{align}
  {\rm D} \phi & = e^{-\phi} {\rm D} e^{\phi} \,, \qquad
  {\rm D} \varphi = e^{-\varphi} {\rm D} e^{\varphi} \,, \qquad
  G_1 = {\rm D} \chi + \tfrac{3}{4} m_4 A^{(1)} \,, \qquad
  G_4 = {\rm D} C + {\vec B}^T \eta {\vec F} \,, \notag \\
  F & = {\rm D} A - \tfrac{1}{2} m_4 B^{(2)} \,, \qquad
  F^{(1)} = {\rm D} A^{(1)} \,, \qquad
  F^{(2)} = {\rm d} A^{(2)} \,, \notag \\
  H^{(1)} & = {\rm D} B^{(1)} - A F^{(1)} -\tfrac{1}{4} m_4 (C - 3 A^{(1)} B^{(2)}) \,, \qquad
  H^{(2)} = {\rm D} B^{(2)} - A F^{(2)} \,.
\label{fieldstrengthsmiia}
\end{align}
The ${\mathbb R}^+$--covariant derivative is defined by
${\rm D}={\rm d} + w_\beta \, m_4 A^{(2)}$ with $w_\beta$ the $\beta$ scaling--weight of the field it acts on, as given in the Table~\ref{9d_weights}, and ${\rm DD} = w_\beta \, m_4 F^{(2)}$. The covariant derivative of the
supersymmetry parameter has no massive deformation.
The explicit deformations of the fermionic field equations read
\begin{align}
  m_4: &
  \begin{cases}
  X_{m_4} (\psi^\mu) = i m_4 e^{\phi/2-3\varphi/2\sqrt{7}} \gamma^{\mu \nu}
    [-i\tfrac{3}{256} \gamma_\nu \lambda
     -i \tfrac{\sqrt{7}}{256} \gamma_\nu \tilde{\lambda} ] \,, \notag \\
  X_{m_4} (\lambda) = - m_4 e^{\phi/2-3\varphi/2\sqrt{7}} \gamma^\nu
    [\tfrac{3}{4}\psi_\nu
    +\tfrac{2}{9\sqrt{7}} i \gamma_\nu \tilde{\lambda}^* ] \,, \notag \\
  X_{m_4} (\tilde{\lambda}) = - m_4 e^{\phi/2-3\varphi/2\sqrt{7}} \gamma^\nu
    [\tfrac{\sqrt{7}}{4}\psi_\nu
    -\tfrac{2}{9\sqrt{7}} i \gamma_\nu \lambda^* ] \,.
  \end{cases}
\end{align}
These massive deformations can be seen as a gauging of the ${\mathbb R}^+$ symmetry $\beta$ with gauge parameter $\beta$ and gauge field transformation
\begin{align}
  \beta: \qquad \Lambda = e^{w_\beta m_4 \beta} \qquad \text{with} \qquad
  A^{(2)} & \rightarrow A^{(2)} - {\rm d} \beta \,.
\label{nonAbtr1}
\end{align}
In addition, we find that the
parabolic subgroup of $SL(2,\mathbb{R})$,
with parameter $\zeta$, is gauged:
\begin{align}
  \zeta:
  \begin{array}{c}
    \chi \rightarrow \chi + \tfrac{3}{4} m_4 \zeta \,, \\
    B^{(1)} \rightarrow B^{(1)} + \tfrac{3}{4} m_4 \zeta B^{(2)} \,,
  \end{array}
  \text{with~~}
  A^{(1)} \rightarrow A^{(1)} - d \zeta + \tfrac{3}{4} m_4 \zeta A^{(2)} \,.
\label{nonAbtr2}
\end{align}
These two scaling symmetries do not commute but rather form the two--dimensional
non-Abelian Lie group A(1), consisting of collinear transformations
\cite{Frankel, Gilmore} (scalings and translations) in one dimension. The algebra reads
\begin{align}
  [ T_\zeta , T_\beta ] = T_\zeta \,,
\end{align}
which is non--semi--simple.

\subsection{Deformations $m_1,m_2,m_3$: SS reduction of IIB using $SL(2,\mathbb{R})$}

We next perform a Scherk-Schwarz reduction of D=10 IIB supergravity
using an Abelian subgroup of the $SL(2,\mathbb{R})$ symmetry. This case has been treated in
\cite{Bergshoeff:1996ui,Lavrinenko:1998qa,Meessen:1998qm,Gheerardyn:2001jj,Hull:2002wg,Bergshoeff:2002mb,Nishino:2002zi}. 
We use the reduction Ans\"atze given in eq.~\eqref{IIBred} with $m_{\text{IIB}}=0$.
This yields massive deformations in D=9
with mass parameters $\vec{m} = (m_1,m_2,m_3)$. Both the explicit and
implicit deformations of the supersymmetry rules can be written in terms of the
superpotential
\begin{align}
  W(\phi, \chi, \varphi) = \ft14 e^{2\varphi/\sqrt{7}}
    \left( m_2 \sinh(\phi) + m_3 \cosh(\phi) + m_1 e^\phi\chi
    - \ft12 (m_2-m_3)e^\phi \chi^2\right ) \,
\end{align}
and the mass matrix employed in the Scherk--Schwarz reduction
\begin{align}
{\cal M} = \ft12
    \left(
    \begin{array}{cc}
      m_1 & m_2 + m_3 \\
      m_2 - m_3 & - m_1
    \end{array} \right) \,.
\end{align}
The explicit deformations are
\begin{align}
  \vec{m}: &
  \begin{cases}
    \delta_{\vec m} \psi_\mu = \ft{1}{7} \gamma_{\mu} W \epsilon \,, \\
    \delta_{\vec m} \lambda = 4i (\frac{\delta W}{\delta \phi}
      +i e^{- \phi} \frac{\delta W}{\delta \chi})\epsilon^* \,, \\
    \delta_{\vec m} \tilde \lambda = 4 i \frac{\delta W}{\delta \varphi} \epsilon^* \,,
  \end{cases}
\label{susyexplmiib}
\end{align}
while the implicit massive deformations read
\begin{align}
  {\rm D} \tau
  & = {\rm d} \tau + 4 e^{-2\varphi/\sqrt{7} -\phi}
    (\frac{\delta W}{\delta \phi} + i e^{-\phi} \frac{\delta W}{\delta \chi}) A \notag \\
  & = \left( {\rm d} + \ft12 [(m_2+m_3) \tau^{-1} + 2 m_1 + (m_3-m_2) \tau] A \right) \tau \,, \notag \\
  F & = {\rm d} A \,, \qquad
  \vec{F} = {\rm d}{\vec A} - \mathcal{M} {\vec B} \,, \qquad
  {\vec H} = {\rm d}{\vec B} - A {\vec F} \,, \notag \\
  G_4 & = {\rm d}C + {\vec B}^{T} \eta {\vec F} + \ft12 {\vec B}^{T} \eta
{\cal M} {\vec B} \,,
\end{align}
for the bosons and
\begin{align}
  D_{\mu} \epsilon & = (\partial_{\mu} + {\omega}_{\mu} + \tfrac{i}{4} e^{\phi} \partial_\mu \chi - i e^{-2 \varphi / \sqrt{7}} W A_\mu ) \epsilon
\end{align}
for the supersymmetry parameter.
The field equations of the 9D fermions receive the following explicit massive corrections:
\begin{align}
  \vec{m}: &
  \begin{cases}
  X_{\vec{m}} (\psi^\mu) = -\gamma^{\mu\nu} [ W \psi_\nu - \tfrac{1}{16} i
    \left(\frac{\delta W}{\delta \phi}+ie^{-\phi}\frac{\delta W}{\delta \chi}\right) \gamma_\nu \lambda^*
    -\tfrac{1}{16} i \frac{\delta W}{\delta \varphi} \gamma_\nu \tilde{\lambda}^* ] \,, \notag \\
  X_{\vec{m}} (\lambda) = - i \gamma^\mu [
    4 \left(\frac{\delta W}{\delta \phi}+ie^{-\phi}\frac{\delta W}{\delta \chi}\right) \psi_\mu^*
    -\tfrac{1}{3} i W \gamma_\mu \lambda
    -\tfrac{8}{9\sqrt{7}} i
    \left(\frac{\delta W}{\delta \phi}+ie^{-\phi}\frac{\delta W}{\delta \chi}\right)
    \gamma_\mu \tilde{\lambda} ] \,, \notag \\
  X_{\vec{m}} (\tilde{\lambda}) = - i \gamma^\nu [
    4 \frac{\delta W}{\delta \varphi} \psi_\nu^* - \tfrac{8}{9\sqrt{7}} i
    \left(\frac{\delta W}{\delta \phi}-ie^{-\phi}\frac{\delta W}{\delta \chi}\right)
    \gamma_\nu \lambda
    - \tfrac{1}{7} i W \gamma_\nu \tilde{\lambda} ] \,.
  \end{cases}
\end{align}

The massive deformations with parameters $\vec{m} = (m_1,m_2,m_3)$ gauge a subgroup of the global $SL(2,\mathbb{R})$ symmetry \eqref{SL2R9D} with parameter and gauge field transformation:
\begin{align}
  SL(2,\mathbb{R}): \qquad \Omega = e^{\mathcal{M} \lambda} \,,
  \text{~~~with~~}
  A \rightarrow A - {\rm d} \lambda \,, \qquad
  \vec{B} \rightarrow \Omega
    ( \vec{B} - \vec{A} \, {\rm d} \lambda ) \,.
\end{align}
Thus these massive deformations correspond to the gauging of the subgroup of $SL(2,\mathbb{R})$ with generator $\mathcal{M}$, the mass matrix employed in the reduction.
Note that the transformations of this subgroup have special properties: for example, the superpotential $W$ is invariant under it. We can distinguish
three distinct cases depending on the value of
$\vec {m}^2 = \tfrac14 (m_1{}^2 + m_2{}^2 -m_3{}^2)$ \cite{Hull:1998vy,Hull:2002wg}:

\begin{itemize}
\item $\vec {m}^2 = 0$.\ \ \ We gauge an ${\mathbb R}$
subgroup of $SL(2,\mathbb{R})$ with parameter $\zeta$ and
invariant metric diag\,(0,1).

\item $\vec {m}^2 > 0$.\ \ \ We gauge an $SO(1,1)^+$
subgroup of $SL(2,\mathbb{R})$ with parameter $\gamma$ and
invariant metric diag\,(1,--1).

\item $\vec {m}^2 < 0$.\ \ \ We gauge an $SO(2)$
subgroup of $SL(2,\mathbb{R})$ with parameter $\theta$ and
invariant metric diag\,(1,1).
\end{itemize}
All these three cases are one--parameter massive deformations.

\subsection{Deformation $m_{\text{IIB}}$: SS reduction of IIB using $\hat\delta$}

Next, we perform a Scherk-Schwarz reduction of D=10 IIB supergravity
using the ${\mathbb R}^+$ symmetry $\hat \delta$ of Table~\ref{IIB_weights}. We use the reduction Ans\"atze given in eq.~\eqref{IIBred} with
$m_1=m_2=m_3=0$. This yields a massive deformation
with parameter $m_{\text{IIB}}$. The explicit deformations of the
supersymmetry rules read
\begin{align}
  m_{\text{IIB}}: &
  \begin{cases}
    \delta_{m_{\text{IIB}}}\psi_\mu = - \ft47 i m_{\text{IIB}} e^{2\varphi/\sqrt7}
      \gamma_\mu \epsilon \,, \\
    \delta_{m_{\text{IIB}}}\tilde\lambda = -\ft{4}{\sqrt{7}} m_{\text{IIB}} e^{2\varphi/\sqrt7} \epsilon^* \,.
  \end{cases}
\end{align}
The implicit deformations read
\begin{align}
F &= {\rm d}A \,, \qquad
\vec{F} = {\rm d}\vec{A} - 2 m_{\text{IIB}} \vec{B} \,, \qquad
\vec{H} = {\rm d}\vec{B} - A\vec{F} \,, \nonumber\\
G_4 &= {\rm d}C + {\vec B}^{T} \eta {\vec F} + m_{\text{IIB}} {\vec B}^{T} \eta {\vec B} \,, \qquad
D\varphi = {\rm d}\varphi - \ft{4}{\sqrt7} m_{\text{IIB}} A \,,
\end{align}
for the bosons and
\begin{align}
  D_{\mu} \epsilon & = (\partial_{\mu} + {\omega}_{\mu} + \tfrac{i}{4} e^{\phi} \partial_\mu \chi + \ft{4}{7} m_{\text{IIB}} \Gamma_{\mu} \slashed{A} ) \epsilon
\end{align}
for the supersymmetry parameter.
The explicit deformations of the fermionic field equations read
\begin{align}
  m_{\text{IIB}}: &
  \begin{cases}
  X_{m_{\text{IIB}}} (\psi^\mu) = i m_{\text{IIB}} e^{2\varphi/\sqrt{7}} \gamma^{\mu\nu}
    [4 \psi_\nu-\tfrac{15}{16\sqrt{7}} i \gamma_\nu \tilde{\lambda}^*] \,, \\
  X_{m_{\text{IIB}}} (\lambda) = m_{\text{IIB}} e^{2\varphi/\sqrt{7}} \gamma^{\nu}
    [ \tfrac{4}{9} i \gamma_\nu \lambda ] \,, \\
  X_{m_{\text{IIB}}} (\tilde{\lambda}) = m_{\text{IIB}} e^{2\varphi/\sqrt{7}} \gamma^{\nu}
    [\tfrac{4}{\sqrt{7}} \psi_\nu^* -i \tfrac{4}{7}
\gamma_\nu \tilde{\lambda}] \, .
  \end{cases}
\end{align}
This is a supergravity where the ${\mathbb R}^+$-symmetry $\delta$ has been gauged:
\begin{align}
  \delta: \qquad \Lambda = e^{w_\delta m_{\text{IIB}} \lambda} \qquad \text{with} \qquad
  A \rightarrow A - {\rm d} \lambda \,, \qquad
  \vec{B} \rightarrow e^{2 m_{\text{IIB}}\lambda} (\vec{B} - \vec{A} \, {\rm d} \lambda) \,.
\end{align}

\subsection{Deformation $m_{11}$: KK reduction of IIA with $m_{11}$-deformation}

Finally, one can also consider the
Kaluza-Klein reduction of the D=10 gauged supergravity discussed
in Subsection~\ref{subsection:10d_gauged_sugra} (see also
Figure~\ref{fig:reductions}).
This leads to a D=9 gauged supergravity with the
following explicit deformations
\begin{align}
  m_{11}: &
  \begin{cases}
    \delta_{m_{11}} {\psi}_{\mu}
     = \ft{9}{14} i m_{11} e^{\phi/2-3\varphi/2\sqrt7} \tau \gamma_\mu \epsilon^* \,, \\
    \delta_{m_{11}} \tilde \lambda =
      - \ft{6}{\sqrt7} m_{11} e^{\phi/2-3\varphi/2\sqrt{7}} \tau^* \epsilon \,.
  \end{cases}
\end{align}
The bosonic implicit deformations read
\begin{align}
&{\rm D}\varphi = {\rm d}\varphi -\ft{6}{\sqrt7} m_{11} A^{(1)} \,,\qquad
F = {\rm D}A + 3 m_{11} B^{(1)} \,,\qquad G_4 = {\rm D}C + \vec{B}^T \eta \vec{F} \,,\nonumber\\
&H^{(1)} = {\rm D}B^{(1)} - A F^{(1)} \,, \qquad
H^{(2)} = {\rm D}B^{(2)} - A F^{(2)} + 3 m_{11} C \,,
\end{align}
with the ${\mathbb R}^+$--covariant derivative of a field with weight $w$
defined by ${\rm D}={\rm d}-w_\alpha m_{11}A^{(1)}$.
For the supersymmetry parameter we find
\begin{align}
  D_{\mu} \epsilon & = (\partial_{\mu} + {\omega}_{\mu} + \tfrac{i}{4} e^{\phi} \partial_\mu \chi + \ft{9}{14} m_{11} \Gamma_{\mu} \slashed{A}^{(1)} )
\epsilon \,.
\end{align}
The fermionic field equations are deformed by the massive contributions
\begin{align}
  m_{11}: &
  \begin{cases}
  X_{m_{11}} (\psi^\mu) = - i m_{11} e^{\phi/2-3\varphi/2\sqrt{7}} \gamma^{\mu \nu}
    [\tfrac{9}{2} \tau \psi_\nu^* -i\tfrac{9}{32} \tau^* \gamma_\nu \lambda
    + i \tfrac{3}{4\sqrt{7}} \tau \gamma_\nu \tilde{\lambda} ] \,, \\
  X_{m_{11}} (\lambda) = m_{11} e^{\phi/2-3\varphi/2\sqrt{7}}
    \gamma^\nu [ - i \tau \tfrac{\sqrt{7}}{6} \gamma_\nu \tilde{\lambda}^* ] \,, \\
  X_{m_{11}} (\tilde{\lambda}) = m_{11} e^{\phi/2-3\varphi/2\sqrt{7}} \gamma^\nu
    [\tfrac{6}{\sqrt{7}} \tau^* \psi_\nu -\tfrac{11}{6\sqrt{7}} i \tau \gamma_\nu \lambda^*
    +\tfrac{1}{7} i \tau^* \gamma_\nu \tilde{\lambda}^* ] \,.
  \end{cases}
\end{align}
This massive deformation is a gauging of the ${\mathbb R}^+$ symmetry $\alpha$:
\begin{align}
  \alpha: \qquad \Lambda = e^{- w_\alpha m_{11} \lambda^{(1)}} \qquad \text{with} \qquad
  A^{(1)} \rightarrow A^{(1)} - {\rm d} \lambda^{(1)}\, .
\end{align}

This reduction does not lead to a new gauged supergravity.
It differs from the $m_{\rm IIA}$ case in the order of the reductions
from D=11. In case 1 one first performs an ordinary KK reduction and next
a SS reduction on $\hat\alpha$ while in the present case the
order of these reductions is reversed: one first performs a
SS reduction on $\hat {\hat \alpha}$ and next an ordinary KK reduction.
Indeed, the difference is just a field redefinition via S--duality plus a
relabelling of the mass parameters: $m_{11} = m_{\rm IIA}$.
The two mass parameters $(m_{11}, m_{\rm IIA})$ form a doublet
under more general $SL(2,\mathbb{R})$ field redefinitions.

\section{Combining Massive Deformations} \label{section:combinations}

In the previous Section we have constructed seven gauged
supergravities, each containing a single mass parameter.
In this Section we would like to consider
combining the massive deformations discussed in the previous Section.
The resulting theories will have more mass parameters characterizing
the different deformations. However, not all combinations will turn out to be 
consistent with supersymmetry. This inconsistency only appears when turning to
the bosonic field equations: the supersymmetry algebra with a combination of
massive deformations always closes, as can be seen from the following argument.

Suppose one has a supergravity with one massive deformation $m$ and
supersymmetry transformations $\delta_0 + \delta_m$.
In all cases discussed in this paper the massive deformation of the
supersymmetry rules satisfies the following property:
$\delta_m (\text{boson}) =0$.
In other words, only the supersymmetry variations of the fermions receive
massive corrections.
This implies that the issue of the closure of the supersymmetry algebra is
a calculation with $m$-independent parts and parts linear in
$m$ but no parts of higher order in $m$~\footnote{That is, up to cubic order in fermions. We have not checked the
higher--order fermionic terms but, based upon dimensional arguments,
we do not expect that these rule out the possibility of
combining massive deformations.}.
On the one hand $[ \delta(\epsilon_1) , \delta(\epsilon_2) ]$ has no terms
quadratic in $m$ since one of the two $\delta$'s acts on a boson.
On the other hand the supersymmetry algebra closes modulo fermionic field
equations which also have only terms independent of and linear in $m$.
Therefore, given the closure of the massless algebra, the closure of the
massive supersymmetry algebra only requires the cancellation of terms
linear in $m$.

In the previous Sections we have not checked the closure of the massive
supersymmetry algebras since this was guaranteed by the higher-dimensional
origin, i.e.~Scherk-Schwarz reduction of supergravity leads to a gauged
supergravity.
However, the argument of linearity allows us to combine different massive
deformations.
Suppose one has two massive supersymmetry algebras with transformations
$\delta_0 + \delta_{m_a}$ and $\delta_0 + \delta_{m_b}$. Both
supersymmetry algebras close modulo
fermionic field equations with (different) massive deformations.
Then the combined massive algebra with transformation
$\delta_0 + \delta_{m_a}+ \delta_{m_b}$ also closes modulo fermionic field
equations whose massive deformations are given by the
sum of the separate massive deformations linear in $m_a$ and $m_b$.
The closure of the combined algebra is guaranteed by the closure of the
two massive algebras since it requires a cancellation at the linear level.

Under supersymmetry variation of the fermionic field equations,
one in general finds linear {\it and} quadratic deformations of the
 bosonic equations of motion.
In addition to these corrections,
we find that there are also
'non-dynamical' equations posing constraints on the mass parameters.
Solving these equations generically excludes the possibility of combining
massive
deformations by requiring mass parameters to vanish.
At first sight, one might seem surprised that the supersymmetry variation
of the fermionic equations of motion leads to constraints other than
the bosonic field equations. However, one should keep in mind that
the multiplets involved cannot be linearized around a Minkowski
vacuum solution. Therefore, the usual rules for linearized (Minkowski)
multiplets do not apply here.

We find that generically adding massive deformations is possible whenever
the D=10 symmetries, giving rise to the separate massive deformations,
can be combined in D=10 as symmetries of IIA or IIB supergravity only.
The combined D=9 supergravity is then
a gauged supergravity which just follows by performing a
SS reduction on the combined D=10 symmetry.

As a warming-up exercise we will in the first Subsection
discuss the situation in D=10. In the next Subsection we will
review the D=9 situation.

\subsection{Combining Massive Deformations in 10D}

The 10D IIA supergravity theory has two massive deformations parameterized
by $m_{\text{R}}$ and ${m_{11}}$.
Can we combine these two massive deformations?
Based on the linearity argument presented above
one would expect a closed supersymmetry algebra.
The bosonic field equations (with up to quadratic deformations)
can be derived by applying the supersymmetry transformations (with only
linear deformations) to the fermionic field equations (containing
only linear deformations). For simplicity,
we truncate all bosonic fields to zero
except the metric and the dilaton. We thus find
\begin{align}
  & (\delta_0+\delta_{m_{\text{R}}}+\delta_{m_{11}})
  ( {{X}}_0 + {{X}}_{m_{\text{R}}} + {{X}}_{m_{11}} )
(\hat{\psi}^{\hat \mu}) = \notag \\
  & \hspace{1cm} = \tfrac{1}{2} \hat{{\Gamma}}^{\hat{{\nu}}}
\hat{{\epsilon}} \,
  [\hat{{R}}^{\hat{{\mu}}}{}_{\hat{{\nu}}}
  - \tfrac{1}{2} \hat{{R}} \hat{{g}}^{\hat{{\mu}}}{}_{\hat{{\nu}}}
  - \tfrac{1}{2} (\partial^{\hat{{\mu}}} \hat{\phi})
(\partial_{\hat{{\nu}}} \hat{\phi})
  + \tfrac{1}{4} (\partial \hat{\phi})^2
\hat{{g}}^{\hat{{\mu}}}{}_{\hat{{\nu}}}
  + \tfrac{1}{4} m_{\text{R}}^2 e^{5 \hat{\phi}/2}
\hat{{g}}^{\hat{{\mu}}}{}_{\hat{{\nu}}}
  + 36 m_{11}^2 e^{-3 \hat{\phi}/2} \hat{{g}}^{\hat{{\mu}}}{}_{\hat{{\nu}}}] + \notag \\
  & \hspace{1.5cm} + \Gamma_{11} \hat{\epsilon}
  [ 3 m_{11} e^{-3 \hat{\phi}/4} \partial^{\hat{\mu}} \hat{\phi} ]
  + \Gamma_{11} \hat{{\Gamma}}^{\hat{{\mu}}} \hat{\epsilon} \,
  [\tfrac{15}{4} m_{\text{R}} m_{11} e^{\hat{\phi}/2} ] \,, \notag \\
  & (\delta_0+\delta_{m_{\text{R}}}+\delta_{m_{11}})
  ( {X}_0 + {X}_{m_{\text{R}}} + {X}_{m_{11}} ) (\hat \lambda) = \notag \\
  & \hspace{1cm} = \hat{\epsilon} \,
    [ \Box \hat{\phi} - \tfrac{5}{4} m_{\text{R}}^2 e^{5 \hat{\phi}/2} ] +
  \hat{{\Gamma}}^{\hat{{\nu}}} \Gamma_{11} \hat{\epsilon}
  [ 9 m_{11} e^{-3 \hat{\phi}/4} \partial_{\hat{\nu}} \hat{\phi} ]
  + \Gamma_{11} \hat{\epsilon} \,
  [\tfrac{33}{2} m_{\text{R}} m_{11} e^{\hat{\phi}/2} ]\,.
\end{align}
At the right-hand side we find four different structures. Three of them
correspond to the field equations of the metric, dilaton and
RR vector. The vector field equation correspond to the terms linear in
$m_{11}$ and containing $\Gamma_{11}$. They show us that truncating
the RR vector to zero forces us to further truncate the dilaton
to $\phi = c$. More interesting is the fourth structure which is
bilinear in $m_{\rm R}m_{11}$. It leads to the constraint
$m_{\text{R}} m_{11} = 0$. This constraint cannot be a remnant of a
higher-rank form field equation due to its lack of Lorentz indices.
It could only fit in the scalar field equation but the $\Gamma_{11}$
factor prevents this.
It is an extra constraint which does not restrict degrees of freedom but
rather restricts mass parameters.

We conclude that, even though the closure of the algebra is a linear calculation and therefore always works for combinations, the bosonic field equations exclude the possibility of the combination of massive deformations in D=10
dimensions.

\subsection{Combining Massive Deformations in 9D}\label{subsection:combinations}

We next try to combine massive deformations in nine dimensions.
One might hope that, due to the large amount of mass parameters, the bosonic field equations do not exclude all possible combinations, like in D=10.
For the present purposes we will focus on specific terms in the
supersymmetry variations of the fermionic field equations.
In the following $\delta_m$ and $X_m$ are understood to mean the supersymmetry variation and fermionic field equation at linear order
containing the sum of all seven possible massive deformations derived in the
previous Section.
Variation of the fermionic field equations gives, amongst other $\gamma$-structures, the terms
\begin{align}
  (\delta_0 + \delta_m)(X_0 + X_m)(\psi^\mu)
  & \sim i \, \gamma^\mu \epsilon [ \ldots ]
    + \gamma^\mu \epsilon^* [ \ldots ] + i \, \gamma^\mu \epsilon^* [ \ldots ] \,, \notag \\
  (\delta_0 + \delta_m)(X_0 + X_m)(\lambda)
  & \sim \epsilon [ \ldots ] + i \, \epsilon [ \ldots ] \,, \notag \\
  (\delta_0 + \delta_m)(X_0 + X_m)(\tilde \lambda)
  & \sim \epsilon [ \ldots ] + i \, \epsilon [ \ldots ] + \epsilon^* [ \ldots ] \,,
\end{align}
where the $[ \ldots ]$ denote different bosonic real expressions of mass bilinears and scalar factors.
These are the analog of the ten-dimensional expression $[ m_{\text{R}} m_{11} e^{\hat{\phi}/2} ]$ we encountered in the
previous Subsection. They are the sources for possible constraints on the
mass parameters.
Requiring all expressions $[ \ldots ]$ to vanish one is led to the
following possible combinations (with the other mass parameters vanishing):
\begin{itemize}
\item {\bf Case 1} with $\{ m_{\rm IIA}, m_4 \}$: this combination can also be obtained by Scherk-Schwarz reduction of IIA employing a linear combination of the symmetries $\hat \alpha$ and $\hat \beta$, guaranteeing its consistency. It is also a gauging of both this symmetry and (for $m_4 \neq 0$) the parabolic subgroup of $SL(2,\mathbb{R})$ in 9D, giving the non-Abelian gauge group A(1).
\item {\bf Case 2,3,4} with $\{ \vec{m}, m_{\rm IIB} \}$:
as in the case with $m_{\rm IIB}=0$ and only $\vec{m}$ this combination contains three different,
inequivalent cases depending on $\vec{m}^2$ (depending crucially on the fact that $m_{\rm IIB}$ is a singlet under $SL(2,\mathbb{R})$):
\begin{itemize}
\item {\bf Case 2} with $\{ \vec{m}, m_{\rm IIB} \}$ and $\vec {m}^2 = 0$.
\item {\bf Case 3} with $\{ \vec{m}, m_{\rm IIB} \}$ and $\vec {m}^2 > 0$.
\item {\bf Case 4} with $\{ \vec{m}, m_{\rm IIB} \}$ and $\vec {m}^2 < 0$.
\end{itemize}
All these combinations can also be obtained
by Scherk-Schwarz reduction of IIB employing a linear combination of the symmetries
$\hat \delta$ and (one of the subgroups of)
$SL(2,\mathbb{R})$, guaranteeing its consistency.
All cases (assuming that $m_{\rm IIB} \ne 0$) correspond
to the gauging of an Abelian non-compact symmetry in 9D. Only the
special case $\{ \vec {m}^2 <0, m_{\rm IIB} = 0 \}$ corresponds to a
SO(2)--gauging.
\item {\bf Case 5} with $\{ m_4 = - \tfrac{12}{5} m_{\rm IIA}, m_2=m_3 \}$:
this case can be understood as the generalized dimensional reduction of Romans' massive
IIA theory, employing the ${\mathbb R}^+$ symmetry that is not broken by the $m_{\rm R}$ deformations:
$\hat \beta -\tfrac{5}{12} \hat \alpha$. It gauges both this linear combination of ${\mathbb R}^+$'s (for $m_4\neq 0$) and
the parabolic subgroup of $SL(2,\mathbb{R})$ (for $m_3 \neq 0$) in 9D, giving the non-Abelian gauge group A(1).
\end{itemize}
Another solution to the quadratic constraints has parameters $\{ m_{\rm IIA}, m_{11} \}$, but
this combination does not represent a new case. It can be obtained from only $m_{\rm IIA}$
(and thus a truncation of Case 1) via an $SL(2,\mathbb{R})$ field redefinition 
(since they form a doublet). Thus the most general deformations are the five cases given above,
all containing two mass parameters. All five of these are gauged theories and have a
higher--dimensional origin. Both case 1 and case 5 have a non-Abelian gauge group provided
$m_4 \neq 0$.

\section{Solutions} \label{section:solutions}

In the first part of this paper we constructed a variety of gauged
supergravities with 32 supersymmetries. They all have in common
that there is a scalar potential. Our next goal is
to make a systematic search for solutions that are based on this
scalar potential. In the next Subsections we will search for two
types of solutions: (i) 1/2 BPS domain-wall (DW) solutions and (ii)
maximally symmetric solutions with constant scalars, i.e.~de Sitter (dS), Minkowski (Mink) or
anti--de Sitter (AdS) solutions.

\subsection{1/2 BPS DW Solutions}

In our previous paper \cite{Bergshoeff:2002mb} we already made a systematic
search for half-supersymmetric DW solutions of the gauged
supergravities corresponding to the cases 3, 4 and 5. Due to a one-to-one
relationship with 7-branes in D=10 dimensions \cite{Meessen:1998qm}
we could even make a systematic investigation of the quantization
of the mass parameters by using the results of
\cite{DeWolfe:1998eu,DeWolfe:1998pr}.

The goal of this Subsection is to investigate whether the five massively deformed
supergravities we found in Subsection~\ref{subsection:combinations} allow
new half-supersymmetric DW solutions. In other words,
we will derive all 1/2 BPS 7-brane solutions to the
9--dimensional supergravities described in the previous
Sections. This analysis should lead, as a check of our calculations, to at least all
the solutions of \cite{Bergshoeff:2002mb}.
Since we are looking for 1/2 BPS solutions it is convenient to solve the Killing spinor equations, which are obtained by setting the supersymmetry variation of the gravitino and dilatinos to zero.
In this way we solve first order equations instead of second order
equations which we would encounter if
we would solve the field equations directly. For static configurations a solution
to the Killing spinor equation is also a solution to the field equations,
so we don't have to explicitly check that the field equations are satisfied.
The projector\footnote{From a general analysis of the possible projectors in 9 dimensions, i.e.~demanding that the projector squares to itself and that its trace is half of the spinor dimension, in order to yield a 1/2 BPS state, we find that there is a second projector given by $\frac{1}{2}(1\pm i\gamma_t)$. This projector would give a euclidean DW, i.e.~a DW having time as a transverse direction. Note that such a Euclidean DW can never be 1/2 BPS since if there existed a Killing spinor it would square to a Killing vector in the {\em transverse} direction, i.e.~time, which is not an isometry of the euclidean DW.} for a DW is given by $\frac{1}{2}(1\pm \gamma_y)$, where $y$ denotes the transverse direction.
We find that, in order to make a projection
operator in the Killing spinor equations, we are forced to set all
mass parameters to zero except for $\vec m$, which corresponds to cases
3, 4 and 5 of Section~\ref{section:combinations}. This is a consistent combination of masses and
we obtain three classes of domain wall solution which were discussed
in detail in \cite{Bergshoeff:2002mb}.
As it turns out, there are no more half-supersymmetric DW solutions.

To summarize, we find that there are no new codimension-one
1/2 BPS solutions
 to the D=9 supergravity theories we obtained in the previous Sections,
as compared to
the three classes of domain wall solutions given in \cite{Bergshoeff:2002mb}.

\subsection{Solutions with Constant Scalars}

In this Subsection we will consider solutions with all three scalars constant.
This is a consistent truncation in two cases which both have two mass
parameters.
In this truncation one is left with the metric only
satisfying the Einstein equation
with a cosmological term
\begin{align}
  R_{\mu \nu} - \tfrac{1}{2} g_{\mu \nu} R =- \Lambda g_{\mu \nu} \,,
\end{align}
with $\Lambda$ quadratic in the two mass parameters. Depending on the sign
of this term one thus has anti-de Sitter, Minkowski or de Sitter geometry.

We find that solutions with constant scalars are possible in the following massive supergravities:
\begin{itemize}
\item {\bf D=10} with $\{ m_{11} \}$ has $\Lambda = 36 m_{11}{}^2
e^{-3\hat\phi/2}$, which gives rise to de Sitter$_{10}$ \cite{Lavrinenko:1998qa}, breaking all supersymmetry. The D=11 origin of this solution
is Mink$_{11}$ written in a basis where the
$x$--dependence is of the required form \cite{Lavrinenko:1998qa}:

\begin{equation}
{\rm Mink_{11}}:\hskip 1truecm ds^2 = e^{2m_{11}x}\bigl (
-dt^2 + e^{2m_{11}t}dx_9^2 + dx^2\bigr )\, .
\end{equation}
\item {\bf D=9, Case 1} with $\{ m_{\rm IIA}= - \tfrac{2}{3} m_4 \}$ has
$\Lambda = \tfrac{63}{4} m_4{}^2e^{\phi -3\varphi/\sqrt{7}}$, which gives rise to De Sitter$_9$, breaking all supersymmetry. This case follows
from the reduction of ${\rm Mink_{10}}$ by using a combination of IIA scale
symmetries that leave the dilaton invariant (since Minkowski has vanishing dilaton) so that, after reduction,
one is left with a non-trivial geometry only.
\item {\bf D=9, Case 4} with $\{ m_{\rm IIB}, m_3 \}$ has
$\Lambda = 28 m_{\rm IIB}{}^2e^{4\varphi/\sqrt{7}}$, which gives rise to
de Sitter$_9$ for non-vanishing $m_{\rm IIB}$.
This case follows
from the reduction of ${\rm Mink_{10}}$
 by using a combination of IIB scale
symmetries that leave the dilaton invariant.
Note that for vanishing
$m_{\rm IIB}$ this reduces to Mink$_9$, despite the presence of $m_3$ \cite{Gheerardyn:2001jj}.
For either $m_{\rm IIB}$ or $m_3$ non-zero this solution breaks all supersymmetry.
\end{itemize}

\section{Conclusions} \label{section:conclusions}

In this paper we have constructed five different D=9 massive deformations
with 32 supersymmetries, each containing two mass parameters.
All these five theories have a higher--dimensional origin via SS reduction from D=10
dimensions. Furthermore, the massive deformations gauge a global symmetry
of the massless theory. The gauge groups we have obtained are the Abelian groups
$SO(2)$, $SO(1,1)^+$, ${\mathbb R}$, ${\mathbb R}^+$ and the unique two--dimensional
non-Abelian Lie group A(1) of scalings and translations on the real line.

We have analyzed the possibility of combining massive deformations
to obtain more general massive supergravities that are not gauged or do not have a
higher--dimensional origin. Our analysis shows that the only possible combinations
are the five two--parameter deformations, which are all gauged and can be uplifted.
We have not made a systematic search for massive D=9 supergravities that are
not the combination of gaugings and we cannot exclude that there are more
possibilities. This requires a separate calculation. In this context,
it is of interest to point out that examples of massive supergravities like
Romans have been found in lower dimensions, e.g.
\cite{Behrndt:2001ab, Louis:2002ny}.
In these cases the compactification manifolds are such
that the candidate gauge fields are truncated away.

It is intriguing that some of the gauged supergravities we have
constructed result from gauging an ${\mathbb R}^+$ scale symmetry that does not leave the
Lagrangian invariant but scales it with a factor. Apparently, it is
possible to gauge such symmetries at the level of the equations
of motion. It would be interesting to work out the general procedure for
doing this.

We now would like to address the question
of whether the gauged supergravities we constructed
can be interpreted as the leading terms in a low-energy approximation
to (compactified) superstring theory. Let us first discuss the status
of the D=10 gauged supergravity. There exist two ways in the literature
to construct this theory:
\begin{description}
\item{(1)} In \cite{Howe:1998qt} the theory was
constructed by pointing out that the Bianchi identities of D=11
superspace allow a more general solution involving a conformal
spin connection. This more general solution
is equivalent to standard D=11 supergravity for a topologically
trivial spacetime but leads to a new possibility for a nontrivial
spacetime of the form $M_{10} \times S^1$. The reduction over the circle
leads to the D=10 gauged supergravity theory.
\item{(2)} In \cite{Lavrinenko:1998qa} the same D=10 gauged supergravity was
obtained via SS reduction of the standard D=11 supergravity using
the ${\mathbb R}^+$ scale symmetry of the D=11 equations of motion.
\end{description}
In both cases it is not obvious how to extend the reduction procedure
beyond the lowest order approximation. The higher-order derivative terms
which arise as corrections in M-theory seem to break the scale
invariance of the D=11 equations of motion\footnote
{We thank Shamit Kachru and Neil Lambert for a discussion on this point.}.
The symmetry used to reduce is therefore only a symmetry of the lowest order approximation.
Presumably this means that the
more general procedure of \cite{Howe:1998qt} involving the conformal
spin connection also does not work in the presence of higher-order
corrections.

One could try to restore the scale invariance by treating the
D=11 Planck length $\ell_{\rm p}$ or, equivalently, the D=11 Einstein constant
$\kappa$, as a scalar field $\ell_{\rm p}(x)$
and giving it a nontrivial weight under the scale transformations. This can
be done by adding to the D=11 Lagrangian a Lagrange multiplier term of the form
\begin{equation}
  \Delta {\cal L} = \int d^{11}x\ \ell_{\rm p}(x) \, d \Lambda^{(10)} \, ,
\end{equation}
where $\ell_{\rm p}(x)$ is the $x$--dependent D=11 Planck length
 and $\Lambda^{(10)}$
is a 10--form Lagrange multiplier field\footnote
{A similar procedure can be performed at the
level of the Green-Schwarz action of the D=11 supermembrane
\cite{Bergshoeff:1992gq} where the membrane tension is replaced
by a worldvolume 2--form potential. This introduces
a scale symmetry in the Green-Schwarz action.
In fact, one can show that in the formulation
of \cite{Bergshoeff:1992gq} the Green-Schwarz action is invariant under the
same scale transformations that leave the equations of motion
of D=11 supergravity invariant.}.
The problem of the above approach is that,
after SS reduction, one is left with {\it two} Lagrange multiplier
fields. The field equation for one of them implies that the
string parameter $\ell_{\rm s}$
is a constant. The other field equation, however, leads
to the constraint that $ m_{11}\ell_{\rm s} = 0$. Thus, one should take
either $m_{11}=0$ or $\ell_{\rm s} =0$. In the first case there is no
deformation left while in the second case one is forced to consider
string theory in the $\ell_{\rm s}=0$ limit
where no higher-order corrections survive.
Naturally, the scale symmetry survives in this limit.

However, the fact that the gauged 10D supergravity with mass parameter $m_{11}$ does not seem to have a higher dimensional origin in the presence of higher derivative corrections does not exclude a possible r\^ole for it in string theory. In this sense its status is similar to Romans' massive theory which also can not be obtained from 11D supergravity plus corrections. Of course the difference is that Romans' theory has a well understood string theory origin which is lacking for the $m_{11}$ theory.

The same discussion carries over to nine dimensions. The massive deformations split up in two categories: those where only the theory to lowest order in $\alpha'$ has a higher--dimensional origin and those where also the higher--derivative corrections can be obtained from 10D. The latter category can be derived using symmetries that extend to all orders in $\alpha'$.
We have two such symmetries:
\begin{itemize}
\item
The $SL(2,\mathbb{R})$ (or rather its $SL(2,\mathbb{Z})$ subgroup) symmetry of IIB.
Thus the $\vec{m}= (m_1, m_2, m_3)$ deformations correspond to the low--energy limits of
three different sectors of compactified IIB string theory (depending on
$\vec{m}^2 = \tfrac{1}{4} (m_1{}^2 + m_2{}^2 - m_3{}^2)$). In \cite{Bergshoeff:2002mb}
DW solutions were constructed for all three sectors. Of these only the D7--brane
has a well--understood role in IIB string theory.
\item
The linear combination $\tfrac{1}{12} \hat \alpha + \hat \beta$ of
${\mathbb R}^+$--symmetries of IIA. Thus one can define
a massive deformation $m_s$ within Case I with
$\{ m_{\rm IIA} = \tfrac{1}{12} m_s, m_4= m_s \}$ which corresponds to
the low--energy limit of a sector of compactified IIA string theory. No vacuum solution
has been constructed for this sector. It would be very interesting to try to find a
vacuum solution and understand which role it plays in IIA string theory.
\end{itemize}

In fact, one can have a better understanding of the $m_s$ massive deformation and the $\tfrac{1}{12} \hat \alpha + \hat \beta$ symmetry of IIA from the following point of view.
The combination $\tfrac{1}{12} \hat \alpha + \hat \beta$ of IIA can be understood from its 11D origin
as the general coordinate transformation $x^{11} \rightarrow \lambda \, x^{11}$. This
explains why all $\alpha'$ corrections transform covariantly under this specific
${\mathbb R}^+$: the higher--order corrections in 11D are invariant under general coordinate
transformations and upon reduction they must transform covariantly under the reduced
g.c.t.'s, among which is the $\tfrac{1}{12} \hat \alpha + \hat \beta$ scaling--symmetry.

The transformation $x^{11} \rightarrow \lambda \, x^{11}$
can also be used for a Scherk--Schwarz reduction from 11D to 9D with a
different procedure to give internal coordinate dependence to the fields.
Let us call this an SS2 reduction as opposed to the SS1 reduction,
which is the method we have used throughout the paper and which is based on {\it global, internal symmetries} of the higher--dimensional theory.
The SS2 procedure \cite{Scherk:1979zr} instead
uses {\it a symmetry of the compactification manifold} for the reduction
Ansatz\footnote
{It was already noted by Scherk and Schwarz that SS1 reduction with a symmetry that originates from a higher--dimensional g.c.t. is equivalent to the corresponding SS2 reduction. For an example, see \cite{Bergshoeff:1997mg}.}.
The massive deformations resulting from a SS2 reduction
can be expressed in terms of the structure constants of the
corresponding non--Abelian gauge group. Using the transformation
$x^{11} \rightarrow \lambda \, x^{11}$
in the SS2 reduction from 11D to 9D we obtain massive deformations
which are equal to the $m_s$ deformations upon relating the components of $f_{ab}{}^c$
to $m_s$. Indeed, this explains why the $m_s$ deformations correspond to a gauging of
the 2D non--Abelian Lie group A(1) rather than only the ${\mathbb R}^+$ symmetry
$\tfrac{1}{12} \alpha + \beta$.

The understanding of the $m_s$ deformation
in terms of a SS2 reduction employing
$x^{11} \rightarrow \lambda \, x^{11}$ also explains why $\tilde{m}_4$ cannot be obtained from
a SS1 reduction. Since S-duality interchanges $x^{10}$ and $x^{11}$, it is the g.c.t.
$x^{10} \rightarrow \lambda \, x^{10}$ that would give rise to a
$m_{11} = \tfrac{1}{12} \tilde{m}_4$ deformation.
However, this transformation is not an internal symmetry of 10D IIA supergravity and thus
cannot be exploited in a SS1 reduction. Since $m_{11}$ does have a 10D origin, this implies
that $\tilde{m}_4$ cannot be obtained from 10D IIA.

The D=9 gauged supergravities involving $m_{11}, m_{\rm IIB}$ or
$m_{\rm IIA} \ne \tfrac{1}{12} m_4$ have
the same status as the D=10 gauged supergravity discussed above,
i.e.~these theories are based upon symmetries that are broken by
$\alpha^\prime$--corrections.
Note that all the de Sitter space solutions we found in Section~\ref{section:solutions}
involve either $m_{11}$, $m_{\rm IIB}$ or $m_{\rm IIA} \neq \tfrac{1}{12} m_4$.
It would be interesting to see whether these de Sitter spaces could occur
as the $\ell_s \rightarrow 0$ limit of an exact solution of string theory.

\medskip
\section*{Acknowledgments}

\noindent We are grateful to Jisk Attema, Sergio Ferrara,
Rein Halbersma, Shamit Kachru,
Neil Lambert, Jan Louis,
Tom\'as Ort\'\i n, Jan Pieter van der Schaar, Kelly Stelle and
Paul Townsend for interesting and useful discussions. This
work is supported in part by the European Community's Human Potential
Programme under contract HPRN-CT-2000-00131 Quantum Spacetime, in which
the University of Groningen is associated with the University of Utrecht.
The work of T.d.W. and U.G. is part of the research program of the
``Stichting voor Fundamenteel Onderzoek der Materie'' (FOM). The work of R.L. 
is supported by the Mexico National Council for Science and Technology
(CONACyT) under grant 010085.
 
\newpage

\appendix

\section{Conventions} \label{appendix:conventions}

We use mostly plus signature $(-+\cdots +)$. All metrics are
Einstein-frame metrics. Doubly hatted fields
and indices are eleven-dimensional, singly hatted fields
and indices ten-dimensional while unhatted ones
are nine-dimensional. Greek indices
$\hat \mu,\hat \nu,\hat \rho\ldots$ denote world coordinates and Latin indices
$\hat a,\hat b,\hat c\ldots$ represent tangent spacetime. They are related by the
Vielbeins $\hat e_{\hat \mu}{}^{\hat a}$ and inverse Vielbeins $\hat e_{\hat a}{}^{\hat \mu}$.
Explicit indices $x,y$ are underlined when flat and non-underlined when curved.
When indices are omitted we use form notation.

\section{Scherk--Schwarz Reduction of Dilaton--Gravity}
\label{appendix:toymodel}

In this Appendix we will discuss in detail the most general Scherk--Schwarz reduction of the dilaton--gravity system.

We start with the truncation of 10D IIA and IIB supergravity to the metric and the dilaton.
The Lagrangian reads
\begin{align}
  \hat{\mathcal{L}} = \sqrt{-\hat{g}} [ \hat{R} - \tfrac{1}{2} (\partial \hat{\phi})^2 ] \,,
\end{align}
while the corresponding Euler--Lagrange equations are given by
\begin{align}
  [ \hat{g}^{\hat\mu \hat\nu} ]: \qquad
  & \hat{R}_{\hat\mu \hat\nu} - \tfrac{1}{2} \hat{R} \hat{g}_{\hat\mu \hat\nu}
    - \tfrac{1}{2} \partial_{\hat \mu} \hat \phi \partial_{\hat \nu} \hat \phi
    + \tfrac{1}{4} (\partial \hat{\phi})^2 \hat{g}_{\hat\mu \hat\nu} = 0 \,, \notag \\
  [ \hat\phi ]: \qquad
  & \Box \hat \phi = 0 \,.
\end{align}
This system has two global symmetries: one can either scale the metric
or one can shift the dilaton:
\begin{align}
  \hat{g}_{\hat\mu \hat\nu} \rightarrow e^{2 m_g} \hat{g}_{\hat\mu \hat\nu} \,, \qquad
  \hat\phi \rightarrow \hat\phi + m_\phi \,.
\end{align}
The shift of the dilaton is a symmetry of the Lagrangian.
The scale transformation of the metric is a symmetry of the field equations only;
it scales the Lagrangian.
This will prove an important difference when performing Scherk--Schwarz reductions.
We will show that one has to reduce field equations, rather than the Lagrangian, when performing SS reductions with symmetries of the field equations only.

Using an arbitrary linear combination of the two global symmetries we make the following Ansatz for Scherk--Schwarz reduction over $x$ to nine dimensions:
\begin{align}
  \hat{e}_{\hat{\mu}}\,^{\hat{a}} & =
  e^{m_g x} \left(\begin{array}{cc}
    e^{\sqrt{7}\varphi/28}e_{\mu}\,^{a} & 0 \\
    0 & e^{-\sqrt{7}\varphi/4}
  \end{array} \right) \,, \qquad
  \hat \phi = \phi + m_\phi x \,,
\end{align}
where we have omitted the Kaluza--Klein vector $A_\mu$ for simplicity.
Using this Ansatz the 10D field equations yield the following 9D equations:
\begin{align}
  [\hat{g}^{\mu \nu}]: \qquad
  & {R}_{\mu \nu} - \tfrac{1}{2} {R} {g}_{\mu \nu}
    - \tfrac{1}{2} \partial_{ \mu} \phi \partial_{ \nu} \phi
    + \tfrac{1}{4} (\partial {\phi})^2 {g}_{\mu \nu}
    - \tfrac{1}{2} \partial_{ \mu} \varphi \partial_{ \nu} \varphi
    + \tfrac{1}{4} (\partial {\varphi})^2 {g}_{\mu \nu} + \notag \\
  & + e^{4 \varphi/\sqrt{7}} (\tfrac{1}{4} m_\phi{}^2+28 m_g{}^2)
  g_{\mu\nu} = 0 \,, \notag \\
  [\hat\phi]: \qquad
  & \Box \phi + 8 m_g m_\phi e^{4 \varphi/\sqrt{7}} = 0 \,, \notag \\
  [\hat{g}^{xx}]: \qquad
  & \Box \varphi - \tfrac{2}{\sqrt{7}} m_\phi{}^2 e^{4 \varphi/\sqrt{7}} = 0 \,.
\label{redfieldeqs}
\end{align}
Note that the field equations of the metric and both scalars get bilinear massive deformations.
In addition one has the reduction of the $\hat{g}^{x \mu}$ field equation
\begin{align}
  [\hat{g}^{x \mu}]: \qquad
  & 2 \sqrt{7} m_g \partial_\mu \varphi
    + \tfrac{1}{2} m_\phi \partial_\mu \phi = 0 \,,
\end{align}
which is the equation of motion for the Kaluza--Klein vector $A_\mu$.
Since it is not important for our argument we will not consider this equation and restrict to \eqref{redfieldeqs}.
We will discuss whether this sector of the field equations can be reproduced by a Lagrangian.

If one performs the SS reduction on the 10D Lagrangian, instead of on the field equations, the result reads $\hat{\mathcal{L}} = e^{8 m_g x} \mathcal{L}$ with the 9D Lagrangian given by
\begin{align}
  \mathcal{L} = \sqrt{-g} [ {R} - \tfrac{1}{2} (\partial {\phi})^2
  - \tfrac{1}{2} (\partial {\varphi})^2
  - V(\phi,\varphi) ] \text{~~~with~~~}
  V(\phi,\varphi) = e^{4 \varphi/7} (\tfrac{1}{2} m_\phi{}^2 + 72 m_g{}^2) \,.
\label{redlagr}
\end{align}
The corresponding Euler--Lagrange equations read
\begin{align}
  [{g}^{\mu \nu}]: \qquad
  & {R}_{\mu \nu} - \tfrac{1}{2} {R} {g}_{\mu \nu}
    - \tfrac{1}{2} \partial_{ \mu} \phi \partial_{ \nu} \phi
    + \tfrac{1}{4} (\partial {\phi})^2 {g}_{\mu \nu}
    - \tfrac{1}{2} \partial_{ \mu} \varphi \partial_{ \nu} \varphi
    + \tfrac{1}{4} (\partial {\varphi})^2 {g}_{\mu \nu} + \notag \\
  & + e^{4 \varphi/\sqrt{7}} (\tfrac{1}{4} m_\phi{}^2+36 m_g{}^2)
  g_{\mu\nu} = 0 \,, \notag \\
  [\phi]: \qquad
  & \Box \phi = 0 \,, \notag \\
  [\varphi]: \qquad
  & \Box \varphi - \tfrac{4}{\sqrt{7}} e^{4 \varphi/\sqrt{7}}
    (\tfrac{1}{2} m_\phi{}^2 + 72 m_g{}^2) = 0 \,.
\label{redEL}
\end{align}
These Euler--Lagrange equations only coincide with the reduction of the 10D Euler--Lagrange equations \eqref{redfieldeqs} provided $m_g = 0$.
Thus the application of SS reduction to the Lagrangian does not give the correct answer if the Lagrangian scales:
the Euler-Lagrange equations \eqref{redEL} are not equal to the field equations \eqref{redfieldeqs} for $m_g \neq 0$ \footnote{
  The difference between substitution in the Lagrangian or its field
equations, in a slightly different context, was also discussed in
  \cite{Townsend:1992fa}.}.
In fact, the situation is worse \cite{Lavrinenko:1998qa}: for $m_g \neq 0$ there {\it is} no Lagrangian $\mathcal{L}$ with potential $V(\phi,\varphi)$ whose Euler-Lagrange equations are the correct field equations \eqref{redfieldeqs}.
The metric field equation would require
\begin{align}
  V(\phi,\varphi)= e^{4 \varphi/\sqrt{7}} (\tfrac{1}{2} m_\phi{}^2 + 56 m_g{}^2) \,,
\end{align}
but this is inconsistent with the $\phi$ and $\varphi$ field equations for $m_g \neq 0$.

Thus we conclude that Scherk--Schwarz reduction on the Lagrangian is only legitimate when the exploited symmetry leaves the Lagrangian {\it invariant} rather than covariant.
For symmetries that scale the Lagrangian one has to reduce the field equations.
Including the full field content, such as the Kaluza--Klein vector $A_\mu$, does not change this conclusion.
One could hope to improve the situation by first going to a frame in which the metric is invariant (possible for any $m_\phi, m_g$ with $m_\phi \neq 0$) and then do the SS reduction.
Since this is related by a field redefinition it will not change the essential properties: the higher--dimensional Lagrangian still scales and the lower--dimensional field equations do not have a corresponding Lagrangian.

\section{Supergravities and their Reductions} \label{appendix:Ansatze}

\subsection{D=11 Supergravity}

The supersymmetry transformation rules of $N=1$ eleven-dimensional supergravity read
\begin{align}
\d \hat{\hat e}_{\hat{\hat\mu}}{}^{\hat{\hat a}}
&= \bar{\hat{\hat\e}} \hat{\hat\Gamma}^{\hat{\hat a}} \hat{\hat\psi}_{\hat{\hat\mu}}\, , \nn\\
\d {\hat{\hat C}}_{\hat{\hat\mu}\hat{\hat\nu}\hat{\hat\rho}}
&= -3\, \bar{\hat{\hat{\e}}} \hat{\hat\Gamma}_{[\hat{\hat\mu}\hat{\hat\nu}} \hat{\hat\psi}_{\hat{\hat\rho}]}\, , \nn\\
\d \hat{\hat\psi}_{\hat{\hat\mu}}
&= D_{\hat{\hat \mu}} \hat{\hat\e} + \ft{1}{192}(\hat{\hat\Gamma}^{(4)} \hat{\hat\Gamma}_{\hat{\hat\mu}} - \ft13 \hat{\hat\Gamma}_{\hat{\hat\mu}} \hat{\hat\Gamma}^{(4)}) \hat{\hat G}_{(4)} \hat{\hat \e}\, ,
\label{11Dsusy}
\end{align}
with the field strengths $\hat{\hat G}_{(4)} = {\rm d} \hat{\hat C}$ and
$D_{\hat{\hat \mu}} \hat{\hat\e} = (\partial_{\hat{\hat \mu}} + \hat{\hat \omega}_{\hat{\hat \mu}}) \hat{\hat\e}$.
The 11D fermionic field content consists solely of a 32--component gravitino, whose field equation reads
\begin{align}
  X_0 (\hat{\hat{\psi}}^{\hat{\hat{\mu}}}) \equiv
  \hat{\hat{\Gamma}}^{\hat{\hat{\mu}}\hat{\hat{\nu}}\hat{\hat{\rho}}}
  \hat{\hat{D}}_{\hat{\hat{\nu}}}
  \hat{\hat{\psi}}_{\hat{\hat{\rho}}} = 0 \,,
\end{align}
with
$\hat{\hat{D}}_{\hat{\hat{\nu}}} = \partial_{\hat{\hat{\nu}}} + \hat{\hat{\omega}}_{\hat{\hat{\nu}}}$
and where we have set the three-form equal to zero.
Under supersymmetry this fermionic field equations transforms into
\begin{align}
  \delta_0 X_0 (\hat{\hat{\psi}}^{\hat{\hat{\mu}}}) =
  \tfrac{1}{2} \hat{\hat{\Gamma}}^{\hat{\hat{\nu}}} \hat{\hat{\epsilon}} \,
  [\hat{\hat{R}}^{\hat{\hat{\mu}}}{}_{\hat{\hat{\nu}}}
   - \tfrac{1}{2} \hat{\hat{R}} \hat{\hat{g}}^{\hat{\hat{\mu}}}{}_{\hat{\hat{\nu}}} ] \,,
\end{align}
which implies the bosonic Einstein equation for the metric.

We use the following reduction Ans\"atze
\begin{align}
{\hat{\hat e}}_{\hat{\hat\mu}}{}^{\hat{\hat a}}
& = e^{m_{11} x} \left(
    \begin{array}{cc} e^{-\hat\phi/12} \hat e_{\hat\mu}{}^{\hat a}
      & - e^{2\hat\phi/3} \hat A_{\hat\mu} \nn\\
      0 & e^{2\hat\phi/3}
    \end{array} \right)\,, \nn \nn\\
\hat{\hat\psi}_{\hat a}
& = e^{-m_{11} x/2} e^{\hat\phi/24} [\hat\psi_{\hat a} - \ft{1}{24} \Gamma_{\hat a} \hat\lambda]\,, \nn\\
\hat{\hat\psi}_{\underline x}
& = \ft13 e^{-m_{11} x/2} e^{\hat\phi/24} \Gamma_{\underline x} \hat\lambda\,, \nn\\
\hat{\hat\e}
& = e^{m_{11} x/2} e^{-\hat\phi/24} \hat\e\,, \nn\\
\hat{{\hat C}}_{\hat \mu \hat \nu \hat \rho}
& = e^{3 m_{11} x} \hat C_{\hat \mu \hat \nu \hat \rho}\,, \quad\quad \hat{\hat C}_{\hat \mu \hat \nu x} = - e^{3 m_{11} x} \hat B_{\hat \mu \hat \nu}\,,
\label{11Dred}
\end{align}
to arrive at the IIA susy-rules in ten dimensions.

\subsection{D=10 IIA Supergravity}

The supersymmetry transformation rules of ten-dimensional IIA supergravity read
\begin{align}
  \delta_0 \hat e_{\hat \mu}{}^{\hat a}
  & = \overline{\hat \epsilon} \Gamma^{\hat a}
    {\hat \psi}_{\hat \mu} \,, \notag \\
  \delta_0 {\hat \psi}_{\hat \mu}
  & = \big( D_{\hat \mu}
    +\ft{1}{48} e^{-\hat \phi/2}( \slashed{\hat H} \hat{\Gamma}_{\hat \mu}
    +\ft{1}{2} \hat{\Gamma}_{\hat \mu} \slashed{\hat H}) \Gamma_{\text{11}}
    +\ft{1}{16} e^{3\hat \phi/4}( \slashed{\hat F} \hat{\Gamma}_{\hat \mu}
    -\ft{3}{4} \hat{\Gamma}_{\hat \mu} \slashed{\hat F}) \Gamma_{\text{11}}+ \notag \\
  & \; \; \; +\ft{1}{192} e^{\hat \phi/4}( \slashed{\hat G} \hat{\Gamma}_{\hat \mu}
    -\ft{1}{4} \hat{\Gamma}_{\hat \mu} \slashed{\hat G})
    \big) \hat \epsilon \,, \notag \\
  \delta_0 {\hat B}_{\hat \mu \hat \nu}
  & = 2 e^{\hat \phi/2} \overline{\hat \epsilon} \Gamma_{\text{11}}
    {\hat \Gamma}_{[\hat \mu} ( {\hat \psi}_{\hat \nu]}
    +\ft18 {\hat \Gamma}_{\hat \nu]} \hat \lambda) \,, \notag \\
  \delta_0 {\hat A}_{\hat \mu}
  & = - e^{-3\hat \phi/4} \overline{\hat \epsilon} \Gamma_{\text{11}}
    ( {\hat \psi}_{\hat \mu} -\ft{3}{8} {\hat \Gamma}_{\hat \mu}
    \hat \lambda) \,, \notag \\
  \delta_0 {\hat C}_{\hat \mu \hat \nu \hat \rho}
  & = - 3 e^{-\hat \phi/4} \overline{\hat \epsilon}
    {\hat \Gamma}_{[\hat \mu \hat \nu}
    ( {\hat \psi}_{\hat \rho]} -\ft{1}{24} {\hat \Gamma}_{\hat \rho]}
    \hat \lambda)
    +3 {\hat A}_{[\hat \mu} \delta_0 {\hat B}_{\hat \nu \hat \rho]} \,, \notag \\
  \delta_0 \hat \lambda
  & = \big( \slashed \partial \hat \phi
    + \ft{1}{12} e^{-\hat \phi/2} \slashed{\hat H} \Gamma_{\text{11}}
    + \ft{3}{8} e^{3\hat \phi/4} \slashed{\hat F} \Gamma_{\text{11}}
    + \ft{1}{96} e^{\hat \phi/4} \slashed{\hat G} \big) \hat \epsilon \,,
    \notag \\
  \delta_0 \hat \phi
  & = \ft{1}{2} \overline{\hat \epsilon} \hat \lambda \,,
\label{IIAsusy}
\end{align}
with the following field strengths:
\begin{align}
  {\hat F} & = {\rm d} {\hat A} \,, \qquad
  {\hat H} = {\rm d} {\hat B} \,, \qquad
  {\hat G} = {\rm d} {\hat C} + {\hat A} {\hat H} \,,
\end{align}
and $D_{\hat{\mu}} \hat{\e} = (\partial_{\hat{\mu}} + \hat{\omega}_{\hat{\mu}}) \hat{\e}$.
Upon (massless) reduction with our Ans\"atze the 11D field equation splits up in two field equations for the 10D IIA fermionic field content, a gravitino and a dilatino:
\begin{align}
  X_0 ({\hat{\psi}}^{{\hat{\mu}}}) & \equiv \hat{{\Gamma}}^{\hat{{\mu}}
\hat{{\nu}}\hat{{\rho}}}
  {\hat{D}}_{\hat{{\nu}}} \hat{{\psi}}_{\hat{{\rho}}}
  - \tfrac{1}{8} (\slashed{\partial} \hat{\phi})
  \hat{{\Gamma}}^{\hat{{\mu}}} \hat{\lambda} = 0 \,, \notag \\
  X_0 (\hat{\lambda}) & \equiv \hat{{\Gamma}}^{\hat{{\nu}}}
 {\hat{D}}_{\hat{{\nu}}} \hat{\lambda}
  - \hat{{\Gamma}}^{\hat{{\nu}}} (\slashed{\partial} \hat{\phi})
\hat{{\psi}}_{\hat{{\nu}}} =0 \,,
\label{X0IIA}
\end{align}
with
${\hat{D}}_{\hat{{\nu}}} = (\partial_{\hat{{\nu}}} + \hat{{\omega}}_{\hat{{\nu}}})$
and where we have set the vector, two- and three-form equal to zero.
Under supersymmetry these fermionic field equations transform into
\begin{align}
  \delta_0 X_0 ({\hat{\psi}}^{{\hat{\mu}}}) =
  & \tfrac{1}{2} \hat{{\Gamma}}^{\hat{{\nu}}} \hat{{\epsilon}} \,
  [\hat{{R}}^{\hat{{\mu}}}{}_{\hat{{\nu}}}
  - \tfrac{1}{2} \hat{{R}} \hat{{g}}^{\hat{{\mu}}}{}_{\hat{{\nu}}}
  - \tfrac{1}{2} (\partial^{\hat{{\mu}}} \hat{\phi}) (\partial_{\hat{{\nu}}} \hat{\phi})
  + \tfrac{1}{4} (\partial \hat{\phi})^2 \hat{{g}}^{\hat{{\mu}}}{}_{\hat{{\nu}}}] \,, \notag \\
  \delta_0 X_0 (\hat{\lambda}) =
  & \hat{\epsilon} \, [ \Box \hat{\phi} ]\,,
\end{align}
which imply the usual graviton-dilaton field equations.

We use the following reduction Ansatz with $x$-dependence implied by the
${\mathbb R}^+$-symmetries $\hat \alpha$ and $\hat\beta$, given in
Table~\ref{IIA_weights}:
\begin{align}
  {\hat e}_{\hat \mu}{}^{\hat a} & = e^{9 m_{\text{IIA}}x/8}\left(
    \begin{array}{cc} e^{\phi/16-3\varphi/16\sqrt{7}} e_\mu{}^a
      & - e^{-7\phi/16+3\sqrt{7}\varphi/16} A_{\mu}^{(2)} \\
      0 & e^{-7\phi/16+3\sqrt{7}\varphi/16}
    \end{array} \right) \,, \notag \\
  {\hat \psi}_a & = e^{-9 m_{\text{IIA}}x/16} e^{-\phi/32 + 3\varphi/32\sqrt{7}}
    [\psi_a +\ft{1}{32} \Gamma_a (\lambda-\ft{3}{\sqrt{7}} {\tilde \lambda})] \,, \notag \\
  {\hat \psi}_{\underline x} & =
    -\ft{7}{32} e^{-9 m_{\text{IIA}}x/16} e^{-\phi/32 + 3\varphi/32\sqrt{7}} \Gamma_{\underline{x}}
    (\lambda - \ft{3}{\sqrt{7}} \tilde\lambda ) \,, \notag \\
  {\hat B}_{\mu \nu} & = e^{3 m_{\text{IIA}}x+m_4 x/2}
    (B_{\mu \nu}^{(2)} - 2 A^{(2)}_{[\mu} A_{\nu]}) \,, \notag \\
  {\hat B}_{\mu x} & = - e^{3 m_{\text{IIA}}x+m_4 x/2} A_\mu \,, \notag
 \displaybreak[2] \\
 {\hat A}_\mu & = -e^{-3m_4 x/4}
    (A_\mu^{(1)} - \chi A^{(2)} ) \,, \notag \\
  {\hat A}_x & = - e^{-3m_4x/4} \chi \,, \notag \\
  {\hat C}_{\mu \nu \rho} & = e^{3 m_{\text{IIA}}x-m_4 x/4}
    (C_{\mu \nu \rho} -3 A^{(1)}_{[\mu} B^{(2)}_{\nu\rho]}
    + 3 A^{(2)}_{[\mu} B^{(1)}_{\nu\rho]}
    + 6 A^{(1)}_{[\mu} A^{(2)}_{\nu} A_{\rho]} ) \,, \notag \\
  {\hat C}_{\mu \nu x} & = - e^{3 m_{\text{IIA}}x-m_4 x/4}
    (B_{\mu \nu}^{(1)} - 2 A^{(1)}_{[\mu} A_{\nu]} ) \,, \notag \\
  {\hat \lambda} & = \ft14 e^{-9m_{\text{IIA}}x/16} e^{-\phi/32 + 3\varphi/32\sqrt{7}}
    (3 \lambda + \sqrt{7} \tilde \lambda ) \,, \notag \\
  \hat \epsilon & = e^{9m_{\text{IIA}}x/16} e^{\phi/32 - 3\varphi/32\sqrt{7}} \epsilon \,, \notag \\
  \hat \phi & = \ft14 (3 \phi + \sqrt{7} \varphi) + \left( \tfrac{3}{2}m_{\text{IIA}}+m_4 \right) x \,.
\label{IIAred}
\end{align}

\subsection{D=10 IIB Supergravity}

The supersymmetry transformation rules of ten-dimensional IIB supergravity read (in complex notation)
\begin{align}
\delta {\hat e}_{\hat \mu}\,^{\hat a}
&= \ft12 \hat {\overline{\epsilon}}\,{\hat \Gamma}^{\hat a}{\hat \psi}_{\hat \mu} + \text{h.c.} \,, \nn\\
\delta \hat \psi_{\hat \mu}
&= D_{\hat \mu} \hat \epsilon - \ft{i}{16 \cdot 5!} \slashed{\hat G}^{\text{(5)}} {\hat \Gamma}_{\hat \mu} \hat \epsilon\, \nn\\
& \quad + \ft{i}{16 \cdot 3!} e^{\hat \phi/2} \left({\hat \Gamma}_{\hat \mu}{\hat \Gamma}^{(3)} + 2 {\hat \Gamma}^{(3)}{\hat \Gamma}_{\hat \mu}\right) \left({\hat H}^{(1)} - \hat\tau {\hat H}^{(2)}\right)_{(3)}{\hat \epsilon}^* \,, \nn\\
\delta \hat \lambda
&= - e^{\hat\phi} \slashed \partial \hat \tau {\hat\epsilon}^* - \ft{1}{2 \cdot 3!} e^{\hat\phi/2} {\hat\Gamma}^{(3)} \left( {\hat H}^{(1)} - \hat\tau {\hat H}^{(2)} \right)_{(3)} \hat \epsilon \,, \nn\\
\delta {\hat B}^{(1)}_{\hat\mu \hat\nu}
&= - e^{\hat\phi/2} {\hat \tau}^* \left(\hat{\overline{\epsilon}}^* {\hat\Gamma}_{[\hat\mu} {\hat \psi}_{\hat\nu]} - \ft{i}{8} \hat{\overline{\epsilon}}\,{\hat\Gamma}_{\hat\mu \hat\nu} \hat\lambda \right) + \text{h.c.} \,, \nn\\
\delta {\hat B}^{(2)}_{\hat\mu \hat\nu}
&= - e^{\hat\phi/2}\left({\hat {\overline{\epsilon}}}^*\,{\hat\Gamma}_{[\hat\mu} {\hat\psi}_{\hat\nu]} - \ft{i}{8} \hat {\overline{\epsilon}}\,{\hat\Gamma}_{\hat\mu \hat\nu} \hat\lambda \right) + \text{h.c.} \,, \nn\\
\delta {\hat D}_{\hat\mu \hat\nu \hat\lambda \hat\rho}
&= 2 i \, \hat {\overline{\epsilon}}\,{\hat\Gamma}_{[\hat\mu \hat\nu \hat\lambda} {\hat\psi}_{\hat\rho]} - \ft32 \, \varepsilon_{ij} {\hat B}^{(i)}_{[\hat\mu \hat\nu}\delta {\hat B}^{(j)}_{ \hat\lambda \hat\rho]} +\text{h.c.} \,, \nn\\
\delta \hat\chi
&= -\ft14 e^{-\hat\phi} \hat {\overline {\epsilon}} {\hat\lambda}^* + \text{h.c.} \,, \nn\\
\delta \hat\phi
&= \ft{i}{4}\hat {\overline {\epsilon}} {\hat\lambda}^* + \text{h.c.} \,,
\label{IIBsusy}
\end{align}
with the complex scalar $\hat\tau = \hat\chi + i e^{-\hat\phi}$ and the field strengths
\begin{align}
  \vec{\hat H} &= {\rm d} \vec{\hat B} \,, \qquad
  \hat G = {\rm d} \hat D + \ft12 \vec {\hat B}^T \eta \vec{\hat H} \,, \qquad 
 \eta = \left( \begin{array}{cc} 0 & 1 \\ -1 & 0 \end{array} \right) \,.
\end{align}
The field strength $\hat G$ is subject to a self--duality constraint.
The covariant derivative of the IIB Killing spinor reads
\begin{align}
   D_{\hat \mu} \hat \epsilon &= (\partial_{\hat \mu} + {\hat \omega}_{\hat \mu}
    + \ft{i}{4} e^{\hat \phi} \partial_{\hat \mu} \hat \chi ) \hat \epsilon \,.
\end{align}
When truncating to the metric, scalars and fermions, the massless 10D IIB fermionic field equations read
\begin{align}
  X_0 ({\hat \psi}^{\hat \mu}) & \equiv
  {\hat \Gamma}^{{\hat \mu}{\hat \nu}{\hat \rho}} (\partial_{\hat \nu} + {\hat{\omega}}_{{{{\hat \nu}}}}+
  \tfrac{1}{4} i e^{{\hat \phi}} \partial_{\hat \nu} {\hat \chi}) {\hat \psi}_{\hat \rho}
  + \tfrac{1}{8} e^{\hat \phi} (\slashed{\partial} {\hat \tau}) {\hat \Gamma}^{\hat \mu} {\hat \lambda}^* =0 \,, \notag \\
  X_0 ({\hat \lambda}) & \equiv
  {\hat \Gamma}^{\hat \mu} (\partial_{\hat \mu} + {\hat{\omega}}_{{{{\hat \mu}}}}+ \tfrac{3}{4} i e^{{\hat \phi}} \partial_{\hat \mu} {\hat \chi}) {\hat \lambda}
  + e^{\hat \phi} {\hat \Gamma}^{\hat \mu} (\slashed{\partial} {\hat \tau}) {\hat \psi}_{\hat \mu}^* = 0 \,.
\end{align}

The reduction Ans\"atze we used for reducing the above rules are
\begin{align}
 \hat{e}_{\hat{\mu}}\,^{\hat{a}} & =
  e^{m_{\text{IIB}} x} \left(\begin{array}{cc}
    e^{\sqrt{7}\varphi/28}e_{\mu}\,^{a} & - e^{-\sqrt{7}\varphi/4}A_{\mu} \\
    0 & e^{-\sqrt{7}\varphi/4}
  \end{array} \right) \,, \nn\\
  \hat{\psi}_{a} & =
   e^{-m_{\text{IIB}} x/2} e^{-\sqrt{7}\varphi/56}\left(\frac{c\tau^*+d}{c\tau+d}\right)^{1/4}(\psi_{a}+
   \ft{1}{8\sqrt{7}}\Gamma_{a}\tilde{\lambda}^*) \,, \nn\\
 \hat{\psi}_{\underline{x}} & =
   -\ft{\sqrt{7}}{8} e^{-m_{\text{IIB}} x/2} e^{-\sqrt{7}\varphi/56}\left(\frac{c\tau^*+d}{c\tau+d}\right)^{1/4}
   \Gamma_{\underline{x}}\tilde{\lambda}^* \,, \nn\\
 \hat{\lambda} & = i e^{-m_{\text{IIB}} x/2} e^{-\sqrt{7}\varphi/56}\left(\frac{c\tau^*+d}{c\tau+d}\right)^{3/4}\lambda \,, \nn\\
 \hat{\epsilon} & = e^{m_{\text{IIB}} x/2} e^{\sqrt{7}\varphi/56}\left(\frac{c\tau^*+d}{c\tau+d}\right)^{1/4}\epsilon
  \displaybreak[2] \,, \nn\\
  \hat{\tau} & = \frac{a\tau+b}{c\tau+d} \,, \nn\\
 \vec{\hat B}_{\mu\nu} & = e^{2 m_{\text{IIB}} x} \Omega(x) {\vec B}_{\mu\nu} 
 \,,\hspace{1cm} \vec{\hat B}_{\mu x} = -e^{2 m_{\text{IIB}} x} \Omega(x) {\vec A}_{\mu} \,, \nn\\
 \hat{D}_{\mu \nu \lambda \rho} & = e^{4 m_{\text{IIB}} x} D_{\mu \nu \lambda \rho} \,,\hspace{1.5cm}
 \hat{D}_{\mu \nu \lambda x}= e^{4 m_{\text{IIB}} x} (- C_{\mu \nu \lambda} + \ft32 \vec{A}^T_{[\mu} \eta \vec{B}_{\nu \rho]}) \,,
\label{IIBred}
\end{align}
where we take the $\Omega$ to be $x$-dependent:
\begin{align}
  \Omega(x) = \left(
    \begin{array}{cc}
      \cosh(\alpha x) + \ft{m_1}{2\alpha}\sinh(\alpha x) &
      \ft{1}{2\alpha}(m_2+m_3)\sinh(\alpha x) \\
      \ft{1}{2\alpha}(m_2-m_3)\sinh(\alpha x) &
      \cosh(\alpha x) - \ft{m_1}{2\alpha}\sinh(\alpha x)
    \end{array} \right)\, .
\end{align}
Upon reduction to 9D the self-duality constraint relates $C_{\mu \nu \lambda}$ to $D_{\mu \nu \lambda \rho}$ and can be used to eliminate the latter.

\subsection{D=9 $N=2$ Supergravity}

The unique nine-dimensional $N=2$ supergravity theory has the following supersymmetry transformations:
\begin{align}
\delta_0 e_{\mu}\,^{a}
&= \ft12 \bar{\epsilon} \gamma^{a} \psi_{\mu} + \text{h.c.} \,,\nn\\
\delta_0 \psi_\mu
&= D_\mu \epsilon + \ft{i}{16} e^{-2 \varphi / \sqrt7} \left( \ft57 \gamma_{\mu} \gamma^{(2)} - \gamma^{(2)} \gamma_{\mu} \right) F_{(2)} \epsilon \nn\\
& \quad - \ft{1}{8\cdot 2!} e^{3 \varphi/ {2\sqrt7}} \left( \ft57 \gamma_{\mu}\gamma^{(2)} - \gamma^{(2)}\gamma_{\mu} \right) \, e^{\phi/2} \left( F^{(1)} - \tau F^{(2)} \right)_{(2)} \epsilon^* \notag \nn\\
& \quad + \ft{i}{8\cdot 3!}e^{- \varphi / {2 \sqrt7}} \left( \ft37 \gamma_{\mu}\gamma^{(3)} + \gamma^{(3)}\gamma_\mu \right) e^{\phi/2} \left( H^{(1)} - \tau H^{(2)} \right)_{(3)}\epsilon^* \notag \nn\\
& \quad - \ft{1}{8\cdot 4!}e^{\varphi / \sqrt7} \left( \ft{1}{7} \gamma_{\mu} \gamma^{(4)} - \gamma^{(4)} \gamma_{\mu} \right) G_4 \epsilon \displaybreak[2] \,,\nn\\[2mm]
\delta_0 \tilde\lambda & = i \slashed \partial \varphi \, \epsilon^* - \ft{1}{\sqrt{7}} e^{-2 \varphi/ \sqrt7} \slashed F \epsilon^* - \ft{3i}{2\cdot 2!\sqrt{7}}e^{3\varphi/{2\sqrt7}} e^{\phi/2}\gamma^{(2)}\left( F^{(1)} - \tau^* F^{(2)} \right)_{(2)} \epsilon \notag \\
& \quad + \ft{1}{2\cdot 3!\sqrt{7}}e^{-\varphi/ {2 \sqrt7}}e^{\phi/2}\gamma^{(3)}\left(H^{(1)} - \tau^* H^{(2)} \right)_{(3)}\epsilon \nn\\
& \quad + \ft{i}{4!\sqrt{7}} e^{\varphi/ \sqrt7} \slashed G_4 \epsilon^* \,,\nn\\[2mm]
\delta_0 \lambda
&= i \slashed{\partial} \phi \, \epsilon^* - e^\phi \slashed{\partial} \chi \, \epsilon^* - \ft{i}{2\cdot 2!} e^{3\sqrt7\varphi/14} e^{\phi/2}\gamma^{(2)}\left( F^{(1)} - \tau F^{(2)} \right)_{(2)}\epsilon \nn\\
& \quad - \ft{1}{2\cdot 3!}e^{-\sqrt{7}\varphi/14}e^{\phi/2}\gamma^{(3)}\left(H^{(1)} - \tau H^{(2)} \right)_{(3)}\epsilon \displaybreak[2] \,,\nn\\[2mm]
\delta_0 A_{\mu}
&= \ft{i}{2} e^{2 \varphi / \sqrt7} \bar{\epsilon}(\psi_{\mu}
- \ft{i}{\sqrt{7}}\gamma_{\mu} \tilde \lambda^*) + \mbox{h.c.} \,,\nn\\
\delta_0 A^{(1)}_{\mu}
&= 
-\ft{i}{2} e^{\phi/2} \tau^* e^{-3\varphi /2\sqrt7} \left( \overline{\epsilon}^* \psi_{\mu} + \ft{i}{4} \overline \epsilon \gamma_{\mu } \lambda + \ft{3i}{4\sqrt7} \overline \epsilon^* \gamma_{\mu} {\tilde \lambda}^* \right) + \text{h.c.} \,,\nn\\
\delta_0 A^{(2)}_{\mu}
&= -\ft{i}{2} e^{\phi/2} e^{-3\varphi/2\sqrt7} \left( \overline{\epsilon}^* \psi_{\mu} + \ft{i}{4} \overline \epsilon \gamma_{\mu} \lambda + \ft{3i}{4\sqrt7} \overline \epsilon^* \gamma_{\mu} {\tilde \lambda}^* \right) + \text{h.c.} \displaybreak[2] \,,\nn\\[2mm]
\delta_0 B^{(1)}_{\mu \nu}
&= 
e^{\phi/2} \tau^* e^{\varphi/2\sqrt7} \left( \overline{\epsilon}^* \gamma_{[\mu} \psi_{\nu]} + \ft{i}{8} {\overline{\epsilon}} \gamma_{\mu \nu} \lambda - \ft{i}{8\sqrt7} {\overline{\epsilon}}^* \gamma_{\mu \nu} {\tilde\lambda}^* \right) - A_{[\mu}^{\phantom{()}} \delta_0 A^{(1)}_{\nu]} + \text{h.c.} \,,\nn\\
\delta_0 B^{(2)}_{\mu \nu}
&= e^{\phi/2} e^{\varphi/2\sqrt7} \left( \overline{\epsilon}^* \gamma_{[\mu} \psi_{\nu]} + \ft{i}{8} {\overline{\epsilon}} \gamma_{\mu \nu} \lambda - \ft{i}{8\sqrt7} {\overline{\epsilon}}^* \gamma_{\mu \nu} {\tilde\lambda}^* \right) - A_{[\mu}^{\phantom{()}} \delta_0 A^{(2)}_{\nu]} + \text{h.c.} \,,\nn\\[2mm]
\delta_0 C_{\m\n\l}
&= \ft{3}{2} e^{-\varphi/\sqrt7} \bar{\e} \gamma_{[\m\n} \left( \psi_{\l]} + \ft{i}{6\sqrt7} \gamma_{\l]} \tilde\lambda^* \right) - \ft32 \vec B_{[\m\n}^{\,T} \eta \,\delta_0 \vec A_{\l]}^{\phantom{T}} + \mbox{h.c.} \displaybreak[2] \,,\nn\\
\delta_0\varphi
&= -\ft{i}{4} \bar{\epsilon} \tilde \lambda^* + \text{h.c.} \,,\nn\\
\delta_0\chi
&= \ft{1}{4} e^{-\phi}\overline{\epsilon}\lambda^* + \text{h.c.} \,,\nn\\
\delta_0\phi
&= -\ft{i}{4} \overline{\epsilon}\lambda^* + \text{h.c.} \,,
\label{9Dsusy}
\end{align}
with the complex scalar $\tau = \chi + i e^{-\phi}$.
The field strengths read
\begin{align}
  G_1 = {\rm d} \chi \,, \qquad
  F = {\rm d} A \,, \qquad
  \vec{F} = {\rm d} \vec{A} \,, \qquad
  \vec{H} = {\rm d} \vec{B} - A \vec{F} \,, \qquad
  G_4 = {\rm d} C + \vec{B}^T \eta \vec{F} \,.
\label{fieldstrengths}
\end{align}
The covariant derivative of the Killing spinor reads
\begin{align}
  D_\mu \epsilon & = (\partial_\mu + \omega_\mu +\ft{i}{4} e^\phi \partial_\mu \chi ) \epsilon \,.
\end{align}
When truncating to the metric, scalars and fermions, the massless 9D fermionic field equations read
\begin{align}
  X_0 (\psi^\mu) & \equiv
  \gamma^{\mu\nu\rho} (\partial_\nu + {{\omega}}_{{{\nu}}}+ \tfrac{1}{4} i e^{\phi} \partial_\nu \chi) \psi_\rho
  - \tfrac{1}{8} e^\phi (\slashed{\partial} \tau) \gamma^\mu \lambda^*
  + \tfrac{1}{8} i (\slashed{\partial} \varphi) \gamma^\mu \tilde{\lambda}^* = 0 \,, \notag \\
  X_0 (\lambda) & \equiv
  \gamma^\mu (\partial_\mu + {{\omega}}_{{{\mu}}}+ \tfrac{3}{4} i e^{\phi} \partial_\mu \chi) \lambda
  + e^\phi \gamma^\mu (\slashed{\partial} \tau) \psi_\mu^* = 0 \,, \notag \\
  X_0 (\tilde{\lambda}) & \equiv
  \gamma^\mu (\partial_\mu + {{\omega}}_{{{\mu}}} - \tfrac{1}{4} i
 e^\phi \partial_\mu \chi) \tilde{\lambda}
  - i \gamma^\mu (\slashed{\partial} \varphi) \psi_\mu^* = 0 \,.
\end{align}
Under supersymmetry these yield the variation
\begin{align}
  \delta_0 X_0 (\psi^\mu) & = \tfrac{1}{2} {{\gamma}}^{{{\nu}}} {{\epsilon}}
  \, [ {{R}}^{{{\mu}}}{}_{{{\nu}}}
  - \tfrac{1}{2} {{R}} {{g}}^{{{\mu}}}{}_{{{\nu}}}
  - \tfrac{1}{2} ((\partial^{{{\mu}}} {\phi}) (\partial_{{{\nu}}} {\phi})
  - \tfrac{1}{2} (\partial {\phi})^2 {{g}}^{{{\mu}}}{}_{{{\nu}}}) + \notag \\
  & \hspace{1.5cm} - \tfrac{1}{2} e^{2\phi} ((\partial^{{{\mu}}} {\chi}) (\partial_{{{\nu}}} {\chi})
  - \tfrac{1}{2} (\partial {\chi})^2 {{g}}^{{{\mu}}}{}_{{{\nu}}})
  - \tfrac{1}{2} ((\partial^{{{\mu}}} {\varphi}) (\partial_{{{\nu}}} {\varphi})
  - \tfrac{1}{2} (\partial {\varphi})^2 {{g}}^{{{\mu}}}{}_{{{\nu}}})] \,, \notag \\
  \delta_0 X_0 (\lambda) & = \epsilon^* [ - e^\phi (\Box \chi + 2 (\partial_\mu \phi)(\partial^\mu \chi)) ] +
  i \epsilon^* [ \Box \phi - e^{2\phi}(\partial \chi)^2 ] \,, \notag \\
  \delta_0 X_0 (\tilde{\lambda}) & = i \epsilon^* [ \Box \varphi ] \,,
\end{align}
which are the massless bosonic field equations for the metric and the scalars.

\section{Spinors and $\Gamma$-matrices in Ten and Nine Dimensions} \label{appendix:spinors}

The $\Gamma$-matrices in ten $(\Gamma_{\hat\mu})$ and nine $(\gamma_\mu)$ dimensions can be chosen to satisfy
\begin{equation}
 \Gamma_{\hat\mu}^\dagger=\eta_{\hat\mu\hat\mu}\Gamma_{\hat\mu}
 \qquad {\rm and} \qquad
 \gamma_\mu^\dagger=\eta_{\mu\mu}\gamma_\mu\,,
\end{equation}
respectively. In ten dimensions we can also choose the $\Gamma$-matrices to be real, while in nine dimensions they will be purely imaginary, which implies that
\begin{equation}
 \Gamma_{\hat\mu}^T=\eta_{\hat\mu\hat\mu}\Gamma_{\hat\mu}
 \qquad {\rm and} \qquad
 \gamma_\mu^T=-\eta_{\mu\mu}\gamma_\mu\,.
\end{equation}
In ten dimensions the minimal spinor is a 32 component Majorana-Weyl spinor with 16 (real) degrees of freedom. With the choice
\begin{equation}
  \Gamma_{11} \equiv -\Gamma_{\underline{0} \cdots \underline{9}} \,,\qquad\qquad
\Gamma_{11}=\left(\begin{matrix} \id & \;\;\,0 \\ 0 & -\id \end{matrix}\right),
\end{equation}
we can write a ten-dimensional Majorana-Weyl spinor as being composed of nine-dimensional, 16 component, Majorana-Weyl spinors according to
\begin{equation}
\psi^{MW}_+=\left(\begin{array}{c}\psi_1\\0\end{array}\right)\,,\qquad\psi^{MW}_-=\left(\begin{array}{c}0\\\psi_2\end{array}\right)\,,
\end{equation}
where $\psi_i$ are nine-dimensional Majorana-Weyl spinors and $+$ or $-$ denotes the chirality of the ten-dimensional spinor. The split of an arbitrary ten-dimensional spinor into two Majorana-Weyl spinors of opposite chirality can of course be done without reference to nine dimensions (through the specific choice of $\Gamma_{11}$), but each ten-dimensional Majorana-Weyl spinor will then in general have 32 non-zero components even though it only has 16 degrees of freedom.
In order to reduce to nine dimensions we use
\begin{equation}
\Gamma_{11}=\sigma_3\otimes\id\,,\qquad
\Gamma_{\underline{x}}=\sigma_1\otimes\id\,,\qquad
\Gamma_{a}=\sigma_2\otimes\gamma_a\,,
\end{equation}
where $x$ is the reduction coordinate and the Pauli matrices are defined as
\begin{equation}
\sigma_1=\left(\begin{matrix}0 & 1 \\ 1 & 0\end{matrix}\right)\,,\qquad
\sigma_2=\left(\begin{matrix}0 & -i \\ i & 0\end{matrix}\right)\,,\qquad
\sigma_3=\left(\begin{matrix}1 & 0 \\ 0 & -1\end{matrix}\right)\,.
\end{equation}
As mentioned above the nine dimensional $\gamma$-matrices are purely imaginary. If we work with a reduction of type IIB, where the two spinors have the same chirality, it may be convenient to introduce complex, nine-dimensional, Weyl spinors according to
\begin{alignat}{2}
\psi_c&= \psi_1+i\psi_2\,,\qquad &\lambda_c=\lambda_2+i \lambda_1\,,\\
\epsilon_c&=\epsilon_1+i\epsilon_2\,,\qquad &\tilde\lambda_c=\tilde\lambda_2+i\tilde\lambda_1\,,
\end{alignat}
which in ten-dimensional notation can be written as, e.g.,
\begin{equation}
\psi^{W}_+=\left(\begin{array}{c}\psi_1\\0\end{array}\right)+i\left(\begin{array}{c}\psi_2\\0\end{array}\right)\,.
\end{equation}
If we instead work with a reduction of type IIA the two spinors will have opposite chirality, and can thus be composed into a ten-dimensional Majorana spinor according to
\begin{equation}
\psi^{M}=\left(\begin{array}{c}\psi_1\\0\end{array}\right)+\left(\begin{array}{c}0\\\psi_2\end{array}\right)\,.
\end{equation}

\begin{figure}[tb]
\begin{center}
\epsfig{file=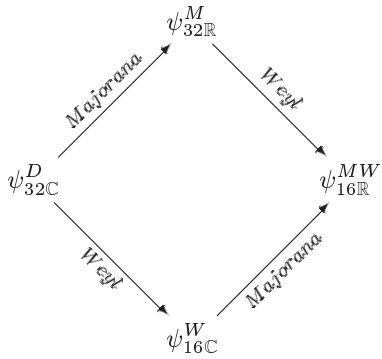}
\caption{Schematic view of how a ten dimensional Dirac spinor can be projected down to a Majorana-Weyl spinor along two different routes. The number of real or complex degrees of freedom for each spinor is also indicated. The relation between the spinors at the intermediate stage (in nine dimensions) is given by (\ref{MWrelation}).}\label{spinordiagram}
\end{center}
\end{figure}

When working with these non-minimal spinors, which are either just Majorana ($\psi_\mu^M$) or just Weyl ($\psi_\mu^W$)~\cite{Bergshoeff:2002mb}, the two formulations are (in nine dimensions) related via
\begin{equation}
\begin{aligned}
  \tfrac{1}{2} (1+\Gamma_{11}) \psi_\mu^M & = \text{Re} (\psi_\mu^W)\, , \\
  \tfrac{1}{2} (1+\Gamma_{11}) \lambda^M & = \text{Im} (\Gamma_{\underline{x}} \lambda^W)\,, \\
  \tfrac{1}{2} (1+\Gamma_{11}) \tilde\lambda^M & = \text{Im} (\Gamma_{\underline{x}} \tilde\lambda^W)\,, \\
  \tfrac{1}{2} (1+\Gamma_{11}) \epsilon^M & = \text{Re} (\epsilon^W)\,, \qquad
\end{aligned}
\begin{aligned}
  \tfrac{1}{2} (1-\Gamma_{11}) \psi_\mu^M & = \text{Im} (\Gamma_{\underline{x}} \psi_\mu^W)\, , \\
  \tfrac{1}{2} (1-\Gamma_{11}) \lambda^M & = \text{Re} (\lambda^W)\, , \\
  \tfrac{1}{2} (1-\Gamma_{11}) \tilde\lambda^M & = \text{Re}
    (\tilde\lambda^W)\, , \\
  \tfrac{1}{2} (1-\Gamma_{11}) \epsilon^M & = \text{Im} (\Gamma_{\underline{x}} \epsilon^W)\, ,
\end{aligned}\label{MWrelation}
\end{equation}
for positive ($\psi_\mu^W, \epsilon^W$) and negative ($\lambda^W,\tilde\lambda^W$) chirality Weyl fermions. With the above mentioned decomposition into nine-dimensional Majorana-Weyl spinors we can write
\begin{equation}
\psi_\mu^M=\left(\begin{array}{c}\psi_1\\\psi_2\end{array}\right)\,,\quad
\epsilon^M=\left(\begin{array}{c}\epsilon_1\\\epsilon_2\end{array}\right)\,,\quad
\lambda^M=\left(\begin{array}{c}\lambda_1\\\lambda_2\end{array}\right)\,,\quad
\tilde\lambda^M=\left(\begin{array}{c}\tilde\lambda_1\\ \tilde\lambda_2\end{array}\right)
\end{equation}
and
\begin{align}
\psi_\mu^W&=\left(\begin{array}{c}\psi_1+i\psi_2\\0\end{array}\right)\,,\quad
\epsilon_\mu^W=\left(\begin{array}{c}\epsilon_1+i\epsilon_2\\0\end{array}\right)\,,\quad\\
\lambda^W&=\left(\begin{array}{c}0\\ \lambda_2+i\lambda_1\end{array}\right)\,,\quad
\tilde\lambda^W=\left(\begin{array}{c}0\\ \tilde\lambda_2+i\tilde\lambda_1\end{array}\right)\,,\quad
\end{align}
where the spinors without an $M$ or $W$ superscript are Majorana-Weyl spinors.
The two different routes to obtain Majorana-Weyl spinors are illustrated in Figure \ref{spinordiagram}.
Note also that it follows from the Clifford algebra and the choice of $\Gamma_{11}$ that $\Gamma_{\underline{x}}$ is off-diagonal, which is crucial for this construction.

\providecommand{\href}[2]{#2}\begingroup\raggedright\endgroup


\end{document}